\documentclass{iopart}
\usepackage{graphicx}
\begin{document}

\title[Directional Locking of Particles Driven Over Periodic Substrates]{Structural Transitions and Dynamical Regimes for Directional 
Locking of Particles Driven Over Periodic Substrates} 
\author{C. Reichhardt  
and C.J. Olson Reichhardt} 
\address{ 
Theoretical Division, 
Los Alamos National Laboratory, Los Alamos, New Mexico 87545 USA}
\ead{charlesr@cnls.lanl.gov}

\begin{abstract}
We numerically investigate collective 
ordering and disordering effects 
for vortices in type-II superconductors 
interacting with square and triangular substrate arrays under 
a dc drive that is slowly rotated
with respect to the fixed substrate. 
A series of directional locking transitions occur as the drive
rotates when
the particle motion locks to 
symmetry directions of the substrate, producing
a series of steps in the velocity-force curves. 
The locking transitions coincide with
structural transitions  
between triangular, square, smectic, or disordered particle arrangements,
which can be identified using the structure factor.
We show that the widths of the locking steps pass through
local minima and maxima as a function of 
the ratio of the number of particles to the number of substrate minima.
Unlike a static system, where 
matching effects occur for simple integer
commensuration ratios, our system exhibits
dynamical commensuration effects where an integer number of particle chains
flow between one-dimensional lines of substrate minima.  As the system
enters and exits the locking steps, order-disorder transitions in the 
structure of the moving particle assembly occur.
We identify two
distinct symmetry locking regimes as a 
function of substrate strength which produce different locking
step characteristics.
For weak substrates, all the particles are in motion 
and a portion of the particles flow through the substrate minima,
leading to structural transitions at certain driving angles. 
For strong substrates, some particles are permanently pinned while the
remaining particles flow around them.
At the crossover between these two regimes of substrate strength, 
some or all of the locking steps are destroyed
due to the onset of chaotic plastic flow which produces
pronounced changes in the transport characteristics.
We show that similar effects occur for colloidal particles
driven over square and triangular substrate arrays.
\end{abstract}
\maketitle

\section{Introduction}

There have been numerous studies of 
collections of driven particles interacting with
periodic substrates in which directional or kinetic locking of the
particle motion occurs
\cite{Nori,Ahn,Korda,Spalding,Lacasta}.
Here, some or all of the particles move along a symmetry direction
of the substrate even though the external driving force is not applied
along the symmetry direction.
This effect was first observed in a superconducting vortex system with 
a square pinning array
in \cite{Nori}. The direction of drive was rotated with
respect to the substrate by applying one dc 
drive $F_D^x$ along a major symmetry axis
of the array while gradually increasing a second perpendicular dc drive 
$F_D^y$.
The initial dc drive
was strong enough to cause all the vortices to flow over the substrate so
that there were no pinned vortices. 
As $F_D^y$ increased, the $y$-component of the response
did not increase smoothly as would be expected in the absence of 
pinning, but instead increased  
in a series of steps which formed as
the vortex motion locked to different symmetry directions of the substrate.
For a square substrate lattice, 
locking steps appear whenever $F^{y}_{D}/F^x_{D} = n/m$,
where $n$ and $m$ are integers, giving a devil's staircase structure. 
Although this initial study was performed for 
superconducting vortices, in \cite{Nori} it was suggested 
that similar effects should occur for other systems of particles 
moving over periodic substrates at varied drive angles,
such as colloidal particles moving over optical substrates.  
A later study predicted similar effects for
the motion of classical charged particles 
over a two dimensional periodic potential array \cite{Ahn}. 
The first experimental observation of this type of symmetry locking 
was obtained for
colloidal particles driven at different angles 
over an optically generated square substrate
\cite{Korda}. The locking, termed kinetic locking in \cite{Korda}, was proposed
as a novel fractionation method for separating different colloidal species
which could be effective whenever one species locked much more strongly
than the other to the symmetry
direction of the optical array.
Such separation
was specifically demonstrated
in other experiments with colloidal particles 
moving over periodic substrates \cite{Spalding}.
Following these works, there have been numerous experiments and simulations 
showing 
symmetry locking for particles moving   
on periodic substrates 
\cite{Lacasta,Olson,AG,Wong,Linke,Reimann,Eichhorn,Drazer1,Drazer,Hermann,Grier,Khoury,Huang,Austin,Hastings,Tierno}, 
as well as new proposals for achieving a similar effect with nanoparticles
\cite{Koplick}. 
Most of these studies 
\cite{Ahn,Korda,Spalding,Lacasta,AG,Wong,Linke,Reimann,Eichhorn,Drazer1,Drazer,Hermann,Grier,Khoury,Huang,Austin,Hastings,Tierno}
involved noninteracting particles and did not consider collective effects. 
In the superconducting system studied in \cite{Nori},
the particle-particle interactions are sufficiently long ranged that
collective particle interactions play an important role in the response
of the system, producing different types of triangular or square
orderings.  This
suggests that there could be a rich variety 
of collective behaviours and even different types of symmetry locking for
particles driven over periodic substrates
that have not been explored yet.  

In addition to the colloidal work, symmetry locking effects have been 
further studied for superconducting vortices
in numerical simulations \cite{Marconi,Carneiro} and have also been observed 
experimentally \cite{Marconi,Surdeanu,Look,Schuller}. 
For systems in which the particles are strongly interacting,
dynamical symmetry locking effects can occur 
even when the particles are moving over a {\it random}
substrate provided that the particle-particle interactions are strong enough
to overcome the randomness of the substrate and 
produce a triangular ordering of the moving particles
\cite{Giamarchi,Olson2}. 
The initial triangular ordering of a lattice of strongly interacting vortices
results in directional locking steps
whenever the direction of the external drive matches one of the lattice
vectors of the triangular vortex lattice.
Under these conditions, the transverse depinning 
threshold for motion perpendicular to the 
lattice symmetry directions is enhanced. 
This effect was predicted in theoretical
work on the moving vortex Bragg glass where the vortex lattice maintains its
triangular ordering \cite{Giamarchi} 
and was also observed in numerical simulations for systems
with weak random substrates \cite{Olson2}. 
The appearance of a critical or nonzero transverse depinning threshold due
to directional locking 
has also been predicted to occur for frictional systems of 
sliding elastic manifolds on periodic surfaces \cite{Achim,Yang,Ying}.  
Vortices and colloids driven over quasicrystalline arrays 
at varied angles also show directional locking effects 
where the motion locks to 
certain orientational degrees of freedom of the lattice 
even though the substrate has no long-range translational
order \cite{OlsonA}.     
    
Since numerical models of vortex and colloidal systems have proved to be
valuable systems for understanding the general behaviour of 
particles moving over periodic 
substrates, 
in this work we use the motion of vortices and colloids to study several aspects
of collective effects in symmetry locking 
that have not been considered previously.
The initial study of the symmetry locking effect for a square substrate
was performed in the limit where all the vortices were
moving \cite{Nori}. 
It would be interesting to understand how the locking effects change 
as the ratio of the external drive to the pinning strength
is varied, particularly when a portion of the particles become pinned for
stronger substrates.
One example of a system exhibiting partial pinning is a superconductor
with an applied magnetic field
of $B/B_{\phi} > 1.0$, where $B_{\phi}$ is the field at which
the number of vortices in the sample equals the number of 
substrate minima in the sample. 
If each substrate minima can trap only one vortex,
then for $B/B_\phi>1$, a portion of the vortices occupy interstitial locations
between the substrate minima.  In order to move under an applied
current, these interstitially confined vortices must either flow along the
occupied rows of pinning by displacing the pinned vortices, or they can flow
through the interstitial regions between the pinning sites
\cite{Harada,Commensurate,Peeters,Reichhardt}. 
Interstitial vortices, when present, are much more mobile than
the pinned vortices
and depin at a lower driving threshold than the vortices at the pinning 
sites \cite{Harada,Reichhardt,Transverse}.
The interstitial vortices may exhibit dynamical locking
effects as the direction of the drive is changed; however, it is also 
possible that
for some driving directions 
the moving interstitial vortices can cause a portion of the vortices at the
pinning sites to depin, disrupting or changing the locking effects.   
The initial symmetry locking study of 
vortices moving through a square pinning array
\cite{Nori} focused only on fields
very near $B/B_{\phi} = 1.0$, and it is not known how the locking
effects would change for higher fields or for varied particle density with a 
fixed pinning density. It might be expected that the 
strength of the locking effects would simply decrease for increasing
particle density; however, in this work we show that 
the locking actually undergoes particle density-induced oscillations related to 
dynamical commensurability effects.   

\section{Simulation}

We model a two-dimensional system with   
periodic boundary conditions. 
For the case of superconducting vortices, 
we consider a sample of size 
$24\lambda \times 24\lambda$, where length is measured in units of the
London penetration depth $\lambda$.
The motion of an individual vortex
evolves according to the overdamped equation 
\begin{equation}
\eta\frac{ d{\bf R}_{i}}{dt} = {\bf F}_{i}^{vv}  + {\bf F}^{p}_{i} + 
{\bf F}^{ext}_{i} .
\end{equation}
Here ${\bf R}_{i}$ is the location of vortex $i$ and
$\eta$ is the damping  
coefficient   
$\eta=\phi_0^2d/2\pi\xi^2\rho_N$,where 
$d$ is the sample thickness, $\xi$ is the superconducting
coherence length, $\rho_N$ is the normal state resistivity,  
and $\phi_0=h/2e$ is the flux quantum.
The first term on the right is the vortex-vortex interaction force
given by
${\bf F}_{i}^{vv} = \sum^{N_{v}}_{j\ne i}f_{0}
K_{1}({R_{ij}}/{\lambda}){\bf {\hat R}}_{ij}$  
where $K_{1}$ is the modified Bessel function, 
$f_{0}=\phi_0^2/(2\pi\mu_0\lambda^3)$, $R_{ij}=|{\bf R}_i-{\bf R}_j|$,
and ${\bf {\hat R}}_{ij}=({\bf R}_i-{\bf R}_j)/R_{ij}$.
Since the Bessel function diverges at the
origin, we place a cutoff on the interaction force at $0.1\lambda$; in 
practice, the vortices do not approach each other this closely. 
The repulsive force falls off exponentially at larger distances 
so we place a maximum interaction distance cutoff
at $6\lambda$ to increase the computational efficiency.   
The periodic substrate is modelled 
as $N_{p}$ parabolic potential traps or pinning sites
of radius $R_{p} = 0.3\lambda$ and maximum strength 
$F_{p}$.  The vortex-substrate interaction has the form
${\bf F}^{p}_{i} =  \sum_{k=1}^{N_p}(F_{p}R_{ik}^{(p)}/R_{p})\Theta((R_{p} - 
R_{ik}^{(p)})/\lambda ){\bf {\hat R}}^{(p)}_{ik}$.
Here $\Theta$ is the Heaviside step function, 
$R_{k}^{(p)}$ is the location of pinning site $k$,
$R^{(p)}_{ik} = |{\bf R}_{i} -{\bf R}^{(p)}_{k}|$, 
and ${\hat {\bf R}}^{(p)}_{ik} =  ({\bf R}_{i} - {\bf R}^{(p)}_{k})/R^{(p)}_{ik}$ 
The pinning sites are placed in a triangular or square arrangement.  

The initial vortex configurations are obtained by
simulated annealing, where we form a high temperature molten state by
applying Langevin kicks to the vortices, and slowly cool the sample to
$T=0$.
We anneal by decreasing $T$ in increments of $\delta T=0.002$ 
with a waiting time of $5000$ simulation time steps between increments.
After annealing, we apply an
external force in the form of a drive applied at an angle that changes
very slowly,
\begin{equation} 
{\bf F}_{ext} = A\sin(\theta(t)){\bf \hat x} + A\cos(\theta(t)){\bf {\hat y}}  
\end{equation}
Here the force amplitude is $A=2.0$,
$\theta=\omega t$, and $\omega$ is the frequency of rotation, which is
taken to be small.
The velocity response in the $x$ and $y$ directions 
can be obtained by measuring the
average vortex velocities 
$\langle V_{x}\rangle = N_{v}^{-1}\sum_{i=1}^{N_{v}}{\bf v}\cdot {\hat {\bf x}}$
and $\langle V_{y}\rangle = N_{v}^{-1}\sum_{i=1}^{N_{v}}{\bf v}\cdot {\hat {\bf y}}$.
We measure the velocity as a function of the drive 
angle $\theta$.
We focus on the case of low frequencies 
to avoid any transient effects in the vortex
dynamics. We have specifically checked that 
the simulated annealing time is long enough that the 
initial vortex configurations do not change for slower annealing rates 
and that the rotation frequency is small
enough that there are no changes in the widths of the
directional locking steps for lower values of $\omega$. 

For the colloidal simulations we use the same type of 
overdamped equation of motion 
and the same form for the pinning substrates. 
The main difference is the
particle-particle interactions, which 
are still repulsive but have
the Yukawa form 
${\bf F}_i^{vv}=-\sum_{j\ne i}^{N_v}\nabla V(R_{ij}){{\bf \hat R}_{ij}}$
with
$V(R_{ij}) = (E_{0}/R_{ij})\exp(-\kappa R_{ij})$, 
where $E_{0} = Z^{*2}/4\pi \epsilon \epsilon_{0}$, 
$\epsilon$ is the solvent dielectric constant, 
$Z^{*}$ is the effective charge of each colloid 
and $1/\kappa$ is the screening length. 
In this case lengths are measured in 
units of $a_{0}$ and forces in units of $F_{0} = E_{0}/a_{0}$.   

\begin{figure}
\includegraphics[width=3.5in]{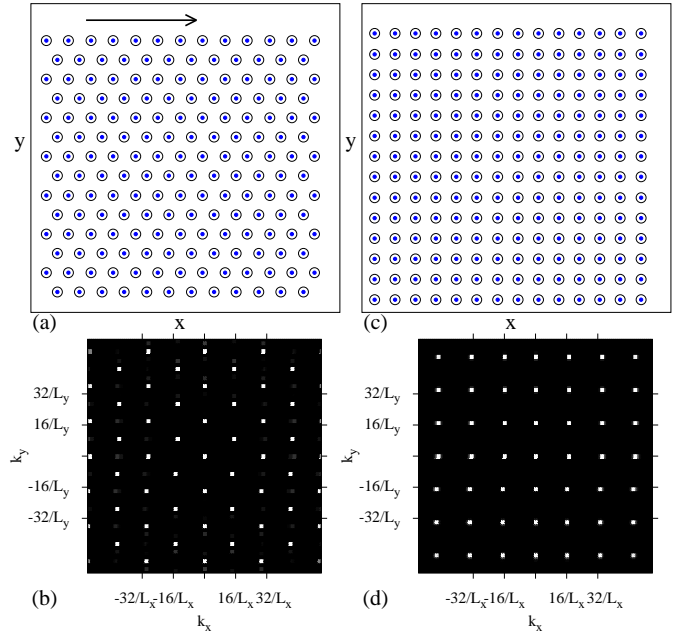}
\caption{
The positions of the pinning sites (large open circles) 
and vortices (small filled circles) 
for a triangular pinning array at $B/B_\phi=1$ and zero drive. 
The arrow indicates the initial direction of the external drive, 
which rotates counterclockwise. 
In this work we consider directional locking for drive angles ranging from 
$\theta = 0^\circ$ to $\theta=90^\circ$. 
(b) The structure factor $S(k)$ for the vortex positions in (a) indicates
triangular ordering. 
(c) The positions of the pinning sites (large open circles) 
and vortex locations (small filled circles) for a square 
pinning array at $B/B_\phi=1$ and zero drive.
(d) $S(k)$ from the vortex positions in (c).  
\label{imagefig}
}
\end{figure}

\section{Structural Transitions and Directional Locking} 

In figure~1(a) we show a snapshot of the system 
for a triangular pinning array at $N_v/N_p=B/B_\phi=1$ and zero drive.
The arrow indicates the initial direction of the applied force, 
which rotates counterclockwise with time.
In figure~1(b) we plot the corresponding structure factor for the
vortex positions,
\begin{equation}
S({\bf k}) = \frac{1}{L^2}\sum_{i,j}\exp(i{\bf k} \cdot[{\bf R}_{i}(t) - {\bf R}_{j}(t)]).
\end{equation}   
Here, six peaks appear which are indicative of the triangular ordering
of the vortex lattice. 
In figure~1(c) we illustrate the nondriven phase for a 
square pinning array at $B/B_{\phi} = 1.0$, along with 
the corresponding structure factor in figure~1(d). 

\begin{figure}
\includegraphics[width=3.5in]{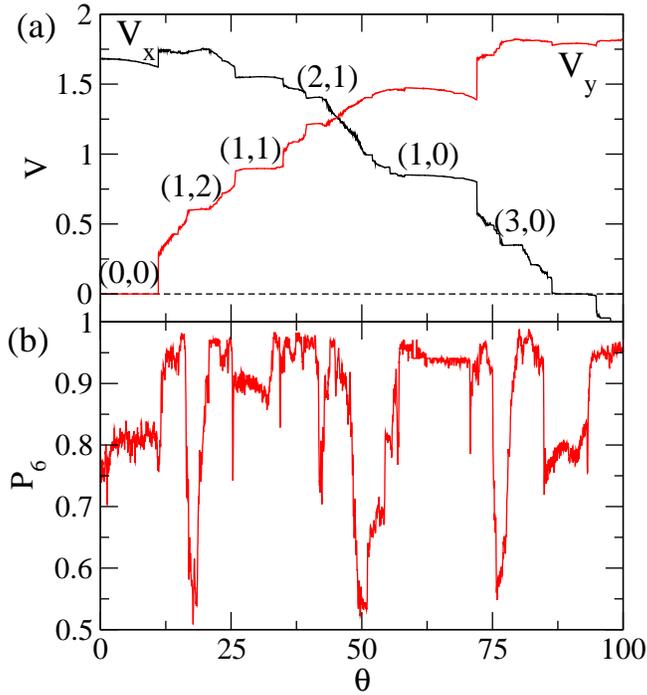}
\caption{
(a) The average velocity $V$ per particle vs 
drive angle $\theta$ for vortices moving on a triangular 
pinning array
with $F_p=1.85$ at $B/B_\phi=1.5$.  
Dark line: $V_x$; light line: $V_{y}$.  Locking steps 
occur at $\theta = \tan^{-1}(\sqrt{3}m/(2n +1))$, where $m,n$ are integers.
Here the locking steps $(m,n)$ at 
(0,0), (1,2), (1,1), (2,1), (1,0), and (3,0)  are highlighted. 
There is also 
a locking step at $\theta=90^\circ$. 
(b) The corresponding fraction of sixfold coordinated particles 
$P_{6}$ vs $\theta$. 
Changes in $P_{6}$ are correlated with the velocity locking steps in (a).     
\label{imagefig2}
}
\end{figure}

We first consider directional locking on a triangular pinning 
substrate, which occurs at driving angles 
$\theta = \tan^{-1}(\sqrt{3}m/(2n +1))$. 
The dominant locking angles fall at
$\theta=60^\circ$ for $m = 1$, $n = 0$ 
and $\theta=30^\circ$ for $m = 1$, $n = 1$. 
We use the notation $(m,n)$ to denote the
locking regions. 
In figure~2(a) we plot $V_{y}$ and $V_{x}$ versus $\theta$
for a sample with $F_{p} = 1.85$ at  $B/B_{\phi} = 1.5$.  
In the absence of pinning, the velocities would follow
a smooth sinusoidal curve with $V_{y}$ starting from 
zero and reaching a maximum at $\theta=90^\circ$ 
when $V_{x}$ goes to zero.   
Figure~2(a) shows that in the presence of pinning, 
both $V_{y}$ and $V_{x}$ pass through a series of pronounced steps
and jumps. For $ \theta < 10^\circ$, 
the motion is locked in the $x$-direction with a finite $V_{x}$
and $V_{y}= 0.0$. 
The value of the $y$ component of the driving force at the end of the
$(0,0)$ step when the locking to the $x$ direction is lost
is the critical transverse depinning force $F^{Tr}_{c}$ 
corresponding to the $(0,0)$ locking. 
At the end of the $(0,0)$ step there is a
discontinuous jump in $V_{y}$ to a finite value which is correlated with
a sudden {\it increase} in $V_{x}$.  The increase in $V_x$ occurs  
even though the $x$ component of the driving force is decreasing.
Over the range $11^\circ < \theta < 16^\circ$, 
$V_{y}$ increases, passing through a small step 
near $\theta=14^\circ$. 
A large locking step is centred at $\theta = 30^\circ$ 
corresponding to the $(1,1)$ locking. 
A second strong step 
in both $V_x$ and $V_y$ occurs at $\theta=60^\circ$ at the
$(1,0)$ locking.
When the system exits this step, $V_{y}$ jumps up 
and $V_{x}$ jumps down. 
In figure~2(b) we plot the fraction of sixfold coordinated particles $P_{6}$
as a function of $\theta$.  Here $P_6=N_v^{-1}\sum_{i=1}^{N_v}\delta(z_i -6)$,
where the coordination number $z_i$ of each particle is obtained from a
Voronoi construction.
In the absence of pinning, $P_6=1.0$ when the vortices form
a triangular lattice.   
On the $(0,0)$ step, 
$P_{6} = 0.8$, indicating that the system 
does not have complete triangular ordering
and that some dislocations must be present. 
Along the $(1,1)$ step, $P_6=0.9$, but $P_{6}$ dips at the start 
and the end of the step.
For the first half of the $(1,2)$ step,  
$P_{6}$ has a low value, and then for the second half
of the step $P_6$ increases.

\begin{figure}
\includegraphics[width=3.5in]{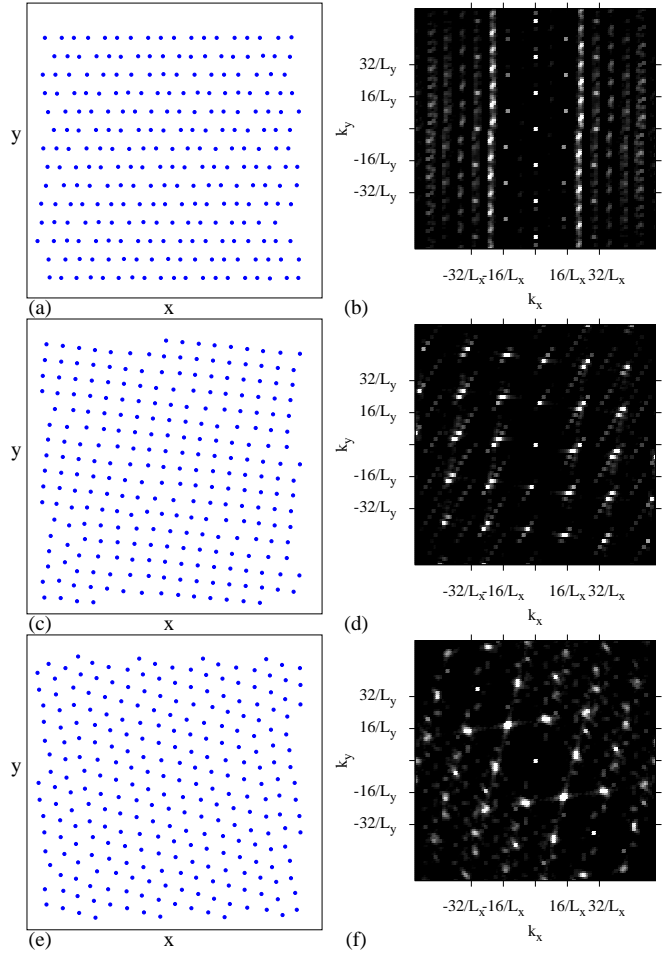}
\caption{
The vortex positions (a,c,e) and corresponding $S(k)$ (b,d,f) for the
triangular pinning
system in figure~2. (a,b) The smectic structure along the (0,0) step.
(c,d) Square lattice formation on the (1,2) step.
(e,f) Partial triangular ordering in a non-step region at $\theta = 36^\circ.$
\label{imagefig3}
}
\end{figure}

To better 
characterize the different vortex structures in the locking regimes,
in figure 3(a) we plot the vortex positions on the $(0,0)$ step for the 
system in figure 2, and in figure 3(b) we show the corresponding $S(k)$. 
Here the particles are moving in one-dimensional channels along the 
pinning rows in the $x$-direction. $S(k)$ has smectic features
with well spaced peaks along the $k_y$-axis indicating that periodic spacing
of the particles along the $y$ direction is induced by the underlying rows
of pinning.  Particles in adjacent rows can slip past each other along
the $x$ direction.
In this regime, $P_{6} = 0.8$ 
due to the formation of aligned dislocations between adjacent rows.
Similar smectic 
ordering has previously been observed for vortices moving over 
random pinning arrays \cite{Moon,GiamarchiA,Pardo,Fangohr} and is associated
with a transverse depinning barrier
\cite{Moon,GiamarchiA,Fangohr2}. 
For vortices moving over pinning arrays with driving 
applied only in the $x$-direction, 
moving smectic phases are 
also possible at finite temperature \cite{Zimanyi,GC}.
In figure 3(a), the smectic ordering occurs 
when all the vortices are confined to move only along the
pinning rows and not between the rows of pinning.  
Since $B/B_{\phi} > 1.0$, this means that the vortex
lattice spacing in the $x$ direction along the pinning rows 
must be smaller than the lattice spacing in the $y$ direction transverse
to the pinning rows, resulting in an effectively anisotropic 
interaction between the vortices. 
Additionally,
certain moving channels may contain more vortices than other 
moving channels, creating dislocations in the vortex lattice.   
Along the $(1,1)$ step, the system has a smectic ordering 
similar to that shown in figure~3(a,b) but tilted by $30^\circ$ with
respect to the $x$ axis.

Figure~3(c,d) illustrates the vortex positions and 
$S(k)$ on the $(1,2)$ locking step. 
In this case the 
vortex lattice has square symmetry
with some disordered regions which produce some smearing in $S(k)$. 
The formation of the square ordering is what 
causes the drop in $P_{6}$ 
on the $(1,2)$ step shown in figure~2(b).
The Voronoi construction we use to determine the particle coordination
numbers $z_i$ can accurately identify triangular ordering but 
is not well suited for measuring slightly disordered square ordering such
as that found on the $(1,2)$ step.
As a result, $P_{6}$ shows large fluctuations on 
the $(1,2)$ step even though the $S(k)$ measurement indicates the presence of
a consistent square ordering throughout the step. 

In figure~3(e,f) we show the vortex positions and $S(k)$ 
for $\theta = 36^\circ$, a non-step region with $P_{6} = 0.95$. 
The system has triangular ordering as indicated by 
the sixfold peaks in $S(k)$. 
Typically the vortex channeling effect is lost in the non-step regions.
For sufficiently weak pinning the vortex-vortex interactions
dominate over the vortex-pin interactions and the vortices form a
mildly disordered triangular lattice, resulting in
large values of $P_{6}$.

Figure~2(a) shows a prominent $(1,0)$ locking step at 
$\theta = 60^\circ$.  Along the step, $V_{y}$ decreases with increasing
$\theta$ until at
$\theta =70.1^\circ$ there is a sharp jump up in 
$V_{y}$ along with a sharp jump down in $V_x$ at the end of the locking step. 
This shows that 
directional locking can induce 
{\it negative differential conductivity}, where the driven particles
actually move more slowly in the direction of drive when the external driving
force is increased.
Negative differential conductivity has been observed
in simulations and experiments for vortices driven in a fixed
direction over periodic pinning arrays 
\cite{Reichhardt,Xiao}. 
The negative differential conductivity effect which we observe here
for triangular pinning did not appear in
previous simulations which considered only
square pinning \cite{Nori}. 
More recently a study with two crossed channels showed that 
a phase locking state could be realized that has velocity-force curves very 
similar to those
in figure~2(a)
and that also exhibits negative differential conductivity \cite{L}. 

\begin{figure}
\includegraphics[width=3.5in]{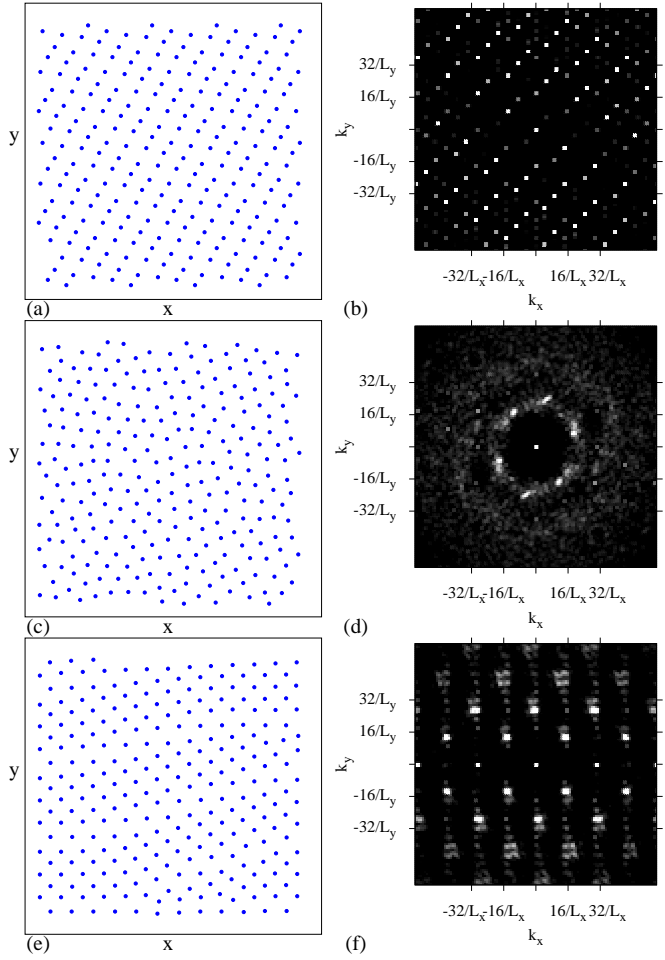}
\caption{
The vortex positions (a,c,e) and corresponding $S(k)$ (b,d,f) for
the triangular pinning system in figure~2. 
(a,b) The smectic structure on the $(1,0)$ step. 
(c,d) An anisotropic liquid at the transition out of the $(1,0)$ step. 
(e,f) A distorted square lattice at the locking regime for 
$\theta=90^\circ$ 
\label{imagefig4}
}
\end{figure}

In figure 2(b), the $(1,0)$ locking step 
has dips in $P_{6}$ at the start and end of the step, while along the
step $P_{6} = 0.945$. 
The dips are caused by the sudden disordering of the particles 
at the transition into and out of the locked regime, and
are associated with a strongly anisotropic vortex lattice structure.
In figure~4(a,b) 
we show the particle positions and 
$S(k)$ for $\theta = 60^\circ$ on the $(1,0)$ step.
The particles again form 
one-dimensional channels and  move along the pinning rows. 
There are some dislocations present in the vortex lattice
which result in smectic ordering; 
however, there is a considerably larger amount of
ordering in the direction transverse to the locking direction than along
the locking direction,
causing the amount of smearing in $S(k)$ to be less than that
observed for the $(0,0)$ step. 
Figure~4(c,d) illustrates 
the system at the dip in $P_{6}$ right at the end of the $(1,0)$ 
locking step.  The
lattice structure is disordered, producing a ringlike structure in $S(k)$. 
Smeared sixfold peaks remain visible in $S(k)$ 
due to the anisotropic nature of the liquidlike structure.

Figure~2(a) shows that the $(3,0)$ locking step 
is accompanied by a strong dip in $P_{6}$. 
Along this step a square vortex lattice forms 
which is similar to the one illustrated for the $(1,2)$ locking step
in figure 3(c,d).  
There is another locking
region centred at $\theta=90^\circ$ where
$V_{x} = 0.0$ and $V_{y}$ shows a cusp feature. 
Along this step $P_{6} = 0.78$, and 
figure~4(e,f) shows that a distorted square lattice appears. 
As $\theta$ increases above $\theta=90^\circ$, the system cycles back
through the same transitions.

\begin{figure}
\includegraphics[width=3.5in]{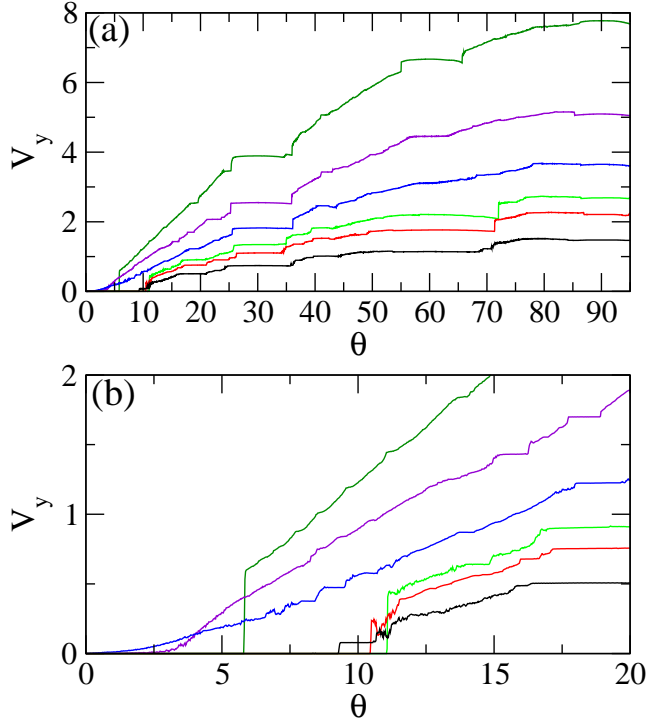}
\caption{
(a) Vortex velocities $V_{y}$ vs $\theta$ for a triangular pinning array with
$F_{p} = 1.85$ 
at $B/B_{\phi} = 0.852$, 1.26, 1.5, 2.0, 2.78, and $4.07$, from bottom right
to top right.
The velocities are normalized by $B_{\phi}$. (b) A blowup of 
panel (a) in the region near the
$(0,0)$ step showing that the 
width of the $(0,0)$ step decreases to zero at $B/B_{\phi}  = 2.0$ and
increases again for $B/B_{\phi} = 4.07$   
\label{Fig5}
}
\end{figure}

\section{Density Dependence and Dynamic Commensuration Effects}

We next consider the effect of particle density on 
the directional locking. 
In figure~5(a) we plot $V_{y}$ for the same triangular 
pinning system from figure 2 with $F_p=1.85$
for different vortex densities $B/B_{\phi} = 0.852$, 1.26, 1.5, 2.0, 2.78, 
and $4.07$. 
As the vortex density increases, the smaller steps become narrower and more
difficult to resolve; 
however, the steps at $(0,0)$, $(1,1)$, $(1,0)$, and $\theta=90^\circ$ remain
clearly visible.
Different steps respond differently to changes in $\theta$.  For example,
for $B/B_{\phi} < 2.0$ the $(0,0)$ and $(1,0)$ steps are present but
at $B/B_{\phi} = 2.0$ these steps are lost. 
In contrast, the step at 
$(1,1)$ is present at all the values of $B/B_\phi$.

The $(0,0)$ and $(1,0)$ steps reappear for 
$B/B_{\phi} \ge 2.78$ and grow in width with increasing vortex density up to 
$B/B_{\phi} = 4.07$.
Figure~5(b) shows a blowup of the region around the $(0,0)$ 
step indicating that at $B/B_{\phi} = 2.0$ the width of the
$(0,0)$ step drops to zero but 
that at $B/B_{\phi} = 4.07$ the step reappears, producing a crossing in 
the $V_y$ versus $\theta$ curves. 
In figure 6 we plot the width $F_c^{Tr}$ of the $(0,0)$ step versus $B/B_\phi$,
where $F_c^{Tr}$ is defined as
$F_c^{Tr}=A\cos(\theta)$, the $y$ component of the force at which the step 
disappears.
For low $B/B_\phi$, $F_c^{Tr}$ is large since 
the system is in the single particle limit when the
vortex-vortex interactions are weak.
As $B/B_{\phi}$ increases, $F_{c}^{Tr}$ deceases
until it reaches a local minimum at $B/B_\phi=0.75$. 
This is followed by a local maximum in $F_c^{Tr}$ at 
$B/B_{\phi} = 1.4$, after which 
$F_{c}^{Tr}$ decreases nearly to zero for $1.8 < B/B_{\phi} < 2.4$.
A broad maximum  in $F_c^{Tr}$ appears
for $3.0 < B/B_{\phi} < 5.5$ before $F_{c}^{Tr}$ 
drops again at higher $B/B_\phi$.  
This nonmonotonic behaviour of $F_c^{Tr}$ contrasts with 
the critical depinning force observed in a system with random pinning,
which monotonically decreases to a saturation level with 
increasing vortex density. 
In recent studies of 
directional locking for vortices moving over quasicrystalline 
pinning arrays \cite{OlsonA},
a nonmonotonic behaviour of the width of the first locking step 
was observed as a function of 
$B/B_{\phi}$. In the quasicrystalline case, the width of the step never
drops completely to zero; however, a broad local minimum occurs
at $B/B_{\phi} =1.6$ and a broad peak appears for
$ 1.6 < B/B_{\phi} < 4.0$, similar to the features shown in figure~6.    

\begin{figure}
\includegraphics[width=3.5in]{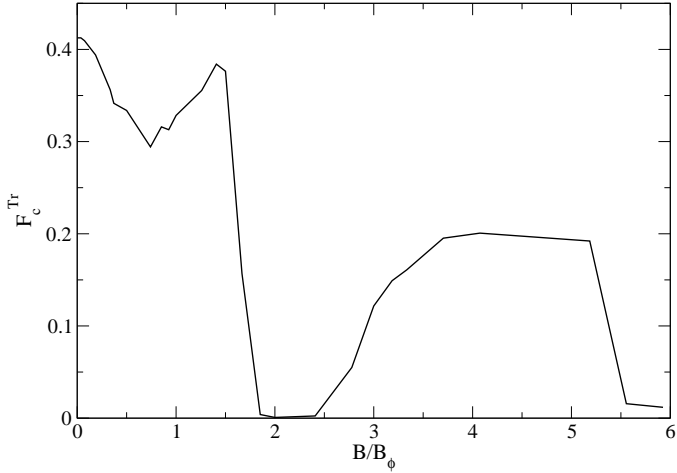}
\caption{
The width $F_c^{Tr}$ of the $(0,0)$ step vs $B/B_{\phi}$ for the system in 
figure 5 with triangular pinning. 
The width is defined as the force component in the $y$-direction, 
$F_{c}^{Tr}=A\cos(\theta)$.
A series of minima and maxima appear that are not simply
related to the commensuration effects 
expected at integer multiples of $B/B_{\phi}$. 
\label{Fig6}
}
\end{figure}

\begin{figure}
\includegraphics[width=3.5in]{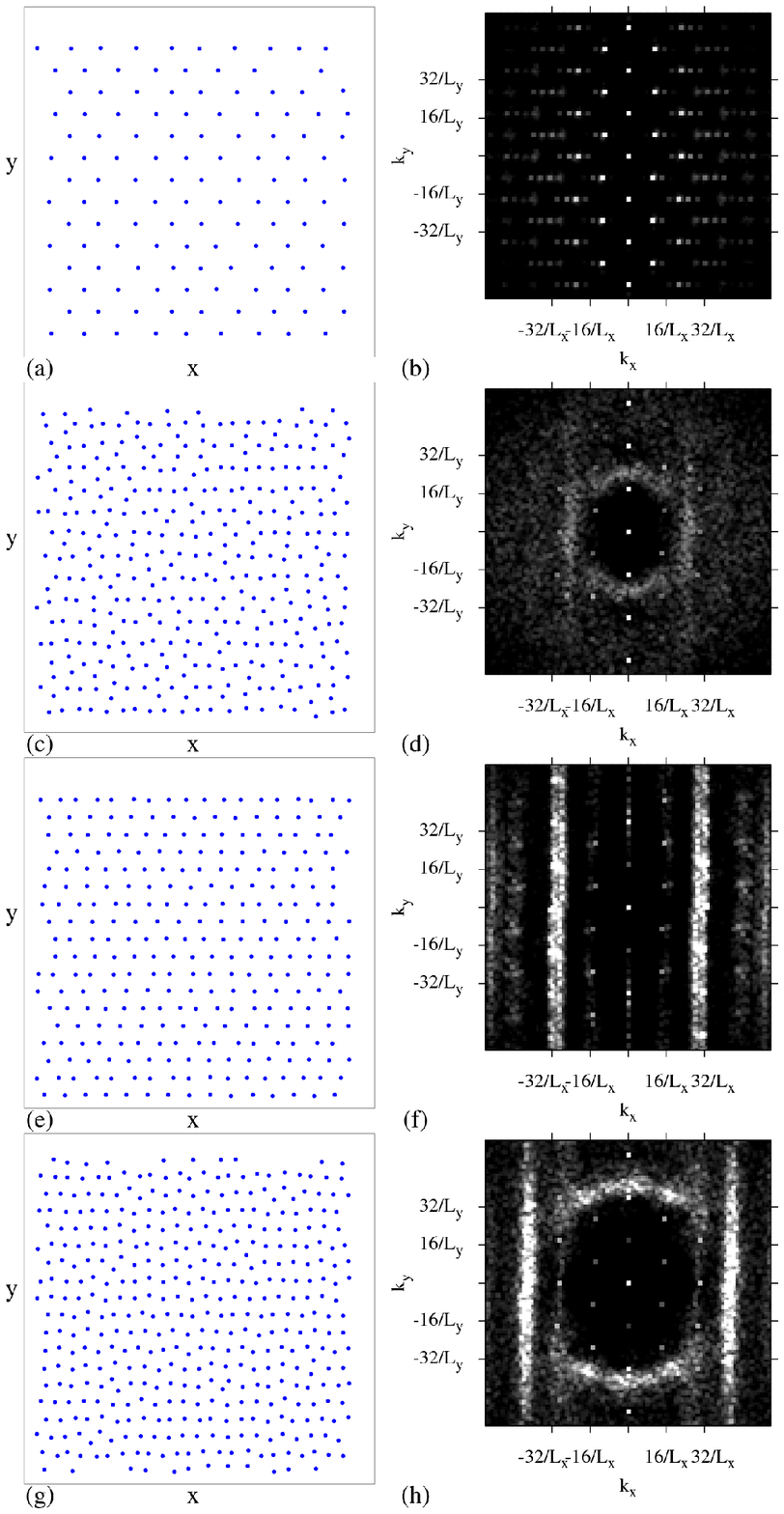}
\caption{
The vortex positions (a,c,e,g) and corresponding $S(k)$ (b,d,f,h) 
for the triangular pinning system from figure 6 with $F_{p} = 1.85$
along the $(0,0)$ step.
(a,b) $B/B_{\phi} = 0.74$. 
(c,d) $B/B_{\phi} = 2.037$. 
(e,f) $B/B_{\phi} = 4.074$. 
(g,h) $B/B_{\phi} = 5.92$. 
\label{Fig.7}
}
\end{figure}

In figure~7 we illustrate the vortex positions and $S(k)$ 
at different vortex densities for the
system in figure~6 along the $(0,0)$ step. 
At $B/B_\phi=0.74$, figure~7(a) shows that a disordered triangular lattice 
forms.
For lower fillings $B/B_\phi<0.74$, the ordering becomes more smectic-like
as the vortex-vortex interactions are reduced in strength.
Smectic ordering also appears close to the peak in $F_c^{Tr}$
at $B/B_{\phi} = 1.5$, as shown in figure~3(a,b).
In the region where $F_c^{Tr}$ drops to zero, the lattice is disordered
and $S(k)$ has a ringlike structure, as shown in 
figure~7(c,d) for $B/B_{\phi} = 2.037$.
There is a tendency for some vortices to form stripe-like structures aligned
with the $x$ direction, generating
weak peaks in $S(k)$ along the $k_y$-axis and making the liquidlike structure
anisotropic.
At $B/B_{\phi} = 4.074$, 
near the middle of the broad peak in $F_{c}^{Tr}$ in figure 6,
figure~7(e,f) indicates that the lattice has 
smectic ordering, while at $B/B_{\phi} = 5.92$ 
where $F_{c}^{Tr}$ drops again,
figure~7(g,h) shows that the lattice is disordered with some partial 
$x$ direction alignment.
For 
$3.8 < B/B_{\phi} < 5.0$ the vortex lattice has smectic structure
while at vortex densities where $F_c^{Tr}$ is low or zero,
the vortex lattice is disordered or nematic.  

In recent work on the directional locking of vortices moving
over quasicrystalline arrays,
for values of $B/B_\phi$ where $F_c^{Tr}$ is enhanced, the vortices have
smectic ordering, while 
for higher $B/B_{\phi}$ where the value of $F_{c}^{Tr}$ drops, 
the system becomes partially disordered \cite{Olson}. 
This is very similar to what we observe for the triangular 
pinning arrays, as shown in figure~6. 
In the quasicrystalline array system, 
square vortex lattices form at fillings where $F_{c}^{Tr}$ passes through a local
minimum but remains finite.
At the local minimum in $F_c^{Tr}$ that we
observe in figure 6 at $B/B_{\phi} = 0.74$, the vortex lattice is not square 
but it does have a distorted triangular ordering, as shown in figure 7(a,b). 
We note that the dynamical ordering which occurs just above a critical
transverse depinning force differs from the dynamical structure that forms
just above the longitudinal depinning force.  In the case of the transverse
depinning, the vortices are already moving and may form a dynamically ordered
configuration even before transverse depinning occurs.  In contrast, for
the longitudinal depinning, the vortices start from a pinned state.
For longitudinal depinning,
peaks in the critical current occur at matching fields 
$B/B_{\phi} = n$ and fractional matching fields $m/n$, 
where $m$ and $n$ are integers \cite{Harada,Commensurate,Peeters,Baret}.
Figure~6 shows that the peaks in the transverse critical current
$F_{c}^{Tr}$ 
do not fall at integer matching fields or fractional matching fields, 
so the local maxima and minima must arise due to some other type
of commensuration effect and not due to simple matching between the number of
vortices and the number of pinning sites. 

\begin{figure}
\includegraphics[width=3.5in]{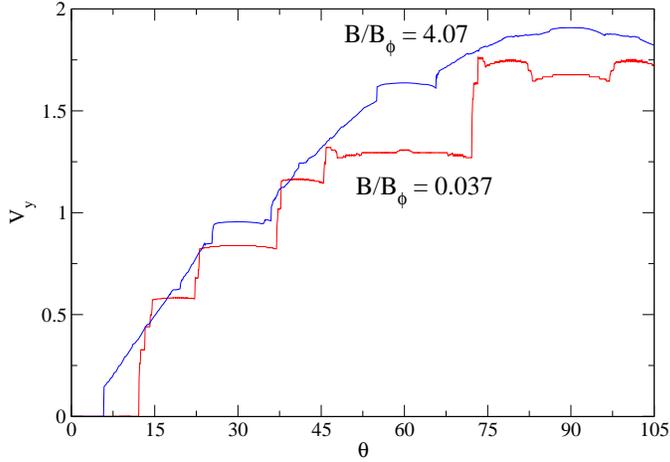}
\caption{
$V_{y}$ vs $\theta$ for the triangular pinning system in figure~5 showing 
a comparison between the low density, single particle behaviour 
limit $B/B_{\phi}=0.037$ (lower line)
and the high density, strongly collective behaviour at $B/B_{\phi} = 4.07$ 
(upper line).
In the noninteracting limit the particles jump directly from one locking
step to another and show no unlocked phases.
\label{Fig.8}
}
\end{figure}

At low vortex densities the system acts in the single particle limit and the
vortices jump from one locking step to another with no unlocked regions
between the steps.
At higher densities, unlocked regions of disordered collective flow appear
between the steps.  A comparison between the low and high density behaviour
appears in figure 8, which shows $V_y$ versus $\theta$ for
$B/B_{\phi}= 0.037$ and 
$B/B_{\phi} = 4.07$.
In the unlocked regions, the vortex-vortex interactions 
are strong enough to permit the vortices to form a triangular lattice.
It is the strong vortex-vortex interactions that destroy 
the higher order steps at the higher densities.

\begin{figure}
\includegraphics[width=3.5in]{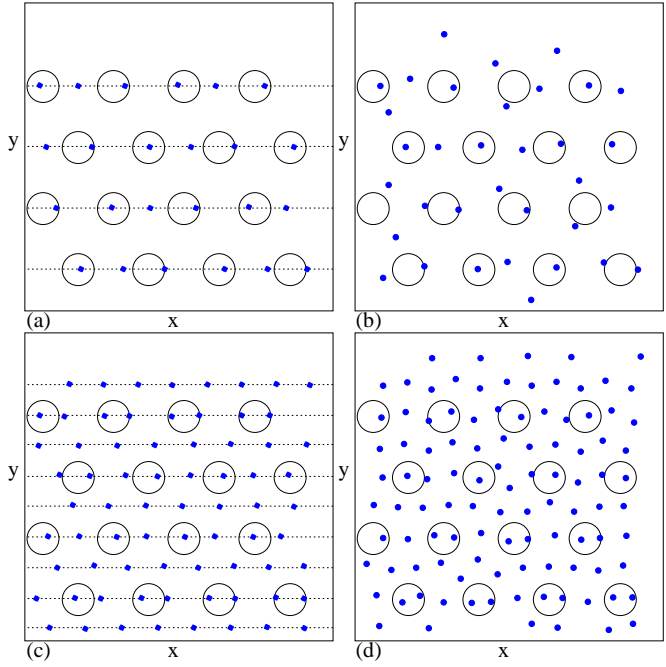}
\caption{
The positions of the pinning sites (large open circles) and vortices (small
filled circles) in a portion of the sample for a triangular pinning array
along the $(0,0)$ step for the system in figure~6. 
(a) $B/B_{\phi} = 1.5$. The dashed lines indicate the motion of the vortices
in one-dimensional channels aligned with the pinning rows.
(b) $B/B_{\phi} = 2.037$, a filling at which $F_c^{Tr}=0$ in figure~6. 
Here the one-dimensional channel structure
shown in panel (a) is lost. 
(c) $B/B_{\phi} = 4.07$, where there are one-dimensional channels of motion
both along and between the pinning sites, as indicated by the dashed
lines.
(d) $B/B_{\phi}  = 5.93$, where $F_c^{Tr}$ is small.
The one-dimensional channel structures are lost.   
\label{Fig.9}
}
\end{figure}

In studies of vortex matter confined to narrow channels, 
the critical current or depinning force
oscillates as a function of vortex density depending on how many
rows of vortices can fit inside the channel \cite{Kes,Karapetrov}.
More recent studies of two-dimensional
periodic pinning arrays show a dynamical commensuration effect 
that occurs in the limit where
the vortices at the pinning sites remain pinned 
but the number of interstitial vortices flowing through
interstitial regions increases with $B/B_\phi$ \cite{Transverse}. 
In this case the transverse depinning force $F^{Tr}_{c}$
exhibits local minima and
maxima as a function of $B/B_{\phi}$. 
At each local maximum, an integer number $n$ of vortex rows 
fit between the pinned vortices and the vortex trajectories are highly ordered. 
As long as this integer number of rows can be maintained,
$F^{Tr}_{c}$ remains high, but eventually a buckling instability
occurs and produces a disordered moving vortex structure which has
a low $F^{Tr}_c$.  As $B/B_\phi$ increases further, a new ordered
vortex state with $n+1$ rows of moving vortices forms and $F^{Tr}_c$
increases again.
The oscillations in $F_{c}^{Tr}$ shown in figure~6 are very similar 
in nature to this effect; however, 
a key difference is that there are no pinned vortices in the
rotating drive system.
Along the $(0,0)$ step, all of the vortices are moving and some of the
vortices are sliding over rows of pinning sites.  The remaining vortices
slide through the interstitial regions between rows of pinning sites.  When
an integer number of interstitial sliding rows fits between adjacent
rows of vortices sliding over the pinning sites, the structure of the
vortex lattice is smectic and there is a local maximum in $F_c^{Tr}$.
At other values of $B/B_\phi$ where an integer number of interstitial sliding
rows is unable to form, the vortex lattice is more disordered and $F_c^{Tr}$ is
low or zero.
We illustrate this effect in figure 9 where we show the locations of the
vortices and pinning sites for different values of $B/B_\phi$ on the 
$(0,0)$ step.
Near the local maximum in $F_c^{Tr}$ at $B/B_\phi=1.5$, 
figure 9(a) indicates that the vortices flow in one-dimensional channels
along the pinning rows and that there are no vortices flowing in the
interstitial regions between rows of pinning.
At $B/B_{\phi} = 2.037$,  figure~9(b) shows that the vortices are no longer
aligned with the pinning rows and that some vortices are flowing through the
interstitial regions.
Near another local maximum in $F_c^{Tr}$  
at $B/B_{\phi} = 4.07$, we show in figure 9(c) that all of the vortices are
flowing in one-dimensional rows and that half of the rows pass through
pinning sites while the other half of the rows pass through interstitial
regions.
At $B/B_{\phi} = 5.93$, near a local minimum in $F_c^{Tr}$, 
figure 9(d) illustrates
that the one-dimensional channel structure is lost.
We expect that for even higher values of $B/B_{\phi}$, additional local
maxima in $F_c^{Tr}$ will occur for fields at which two, three, or higher
integer numbers of rows of vortices can be accommodated in the interstitial
regions.
This same mechanism of the formation of integer numbers of ordered
one-dimensional flowing channels also produces the local minima and maxima
in the transverse critical force 
$F^{Tr}_{c}$ found for quasicrystalline pinning arrays in \cite{OlsonA}.
In the quasicrystalline system, the existence of orientational order is 
sufficient to permit the formation of ordered channels of flow, and 
translational order is not required.
Systems with random pinning substrates exhibit no oscillations
in $F^{Tr}_{c}$ 
as a function of field since these substrates lack long-range
orientational order.
We thus expect that any periodic or quasiperiodic pinning substrate
which has orientational order will produce oscillations in the
transverse depinning force as a function of particle density.  
Similar oscillations of the widths of the higher order locking steps 
with increasing field should also occur; however, since the spacing 
between the one-dimensional channels of particles flowing along the pinning
sites will vary from step to step,
the particle densities at which the local minima or maxima 
in step width occur may be different than the 
densities at which maxima and minima of $F_c^{Tr}$ for the $(0,0)$ step
occur.  

\begin{figure}
\includegraphics[width=3.5in]{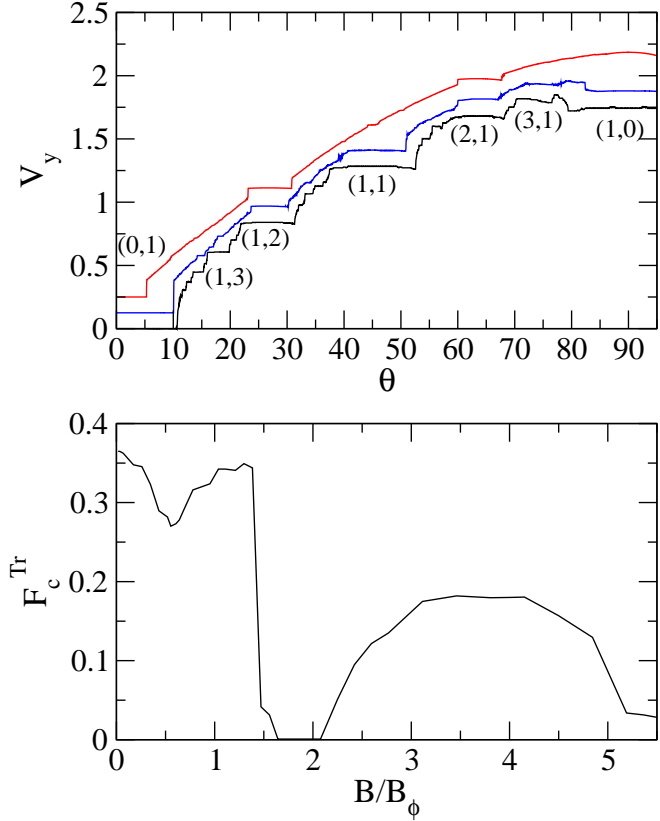}
\caption{
(a) $V_{y}$ vs $\theta$ for a 
sample with a square pinning array at $F_{p} = 1.5$  
for $B/B_{\phi} = 0.2$, $1.5$, and $4.0$ (from bottom to top). 
Locking steps occur at $\theta = \tan^{-1}(m/n)$ where $m$ and $n$ are integers. 
The steps at $(m,n)$= (0,1), (1,3), (1,2), (1,1), (2,1), (3,1), and
(1,0) are marked 
The $B/B_\phi=0.2$ and $B/B_\phi=1.5$ curves have been shifted up for 
clarity. 
(b) The width of the first step given by $F_c^{Tr}$, the value of the force component
in the $y$ direction, vs $B/B_{\phi}$ for the same system. 
\label{Fig10}
}
\end{figure}

\begin{figure}
\includegraphics[width=3.5in]{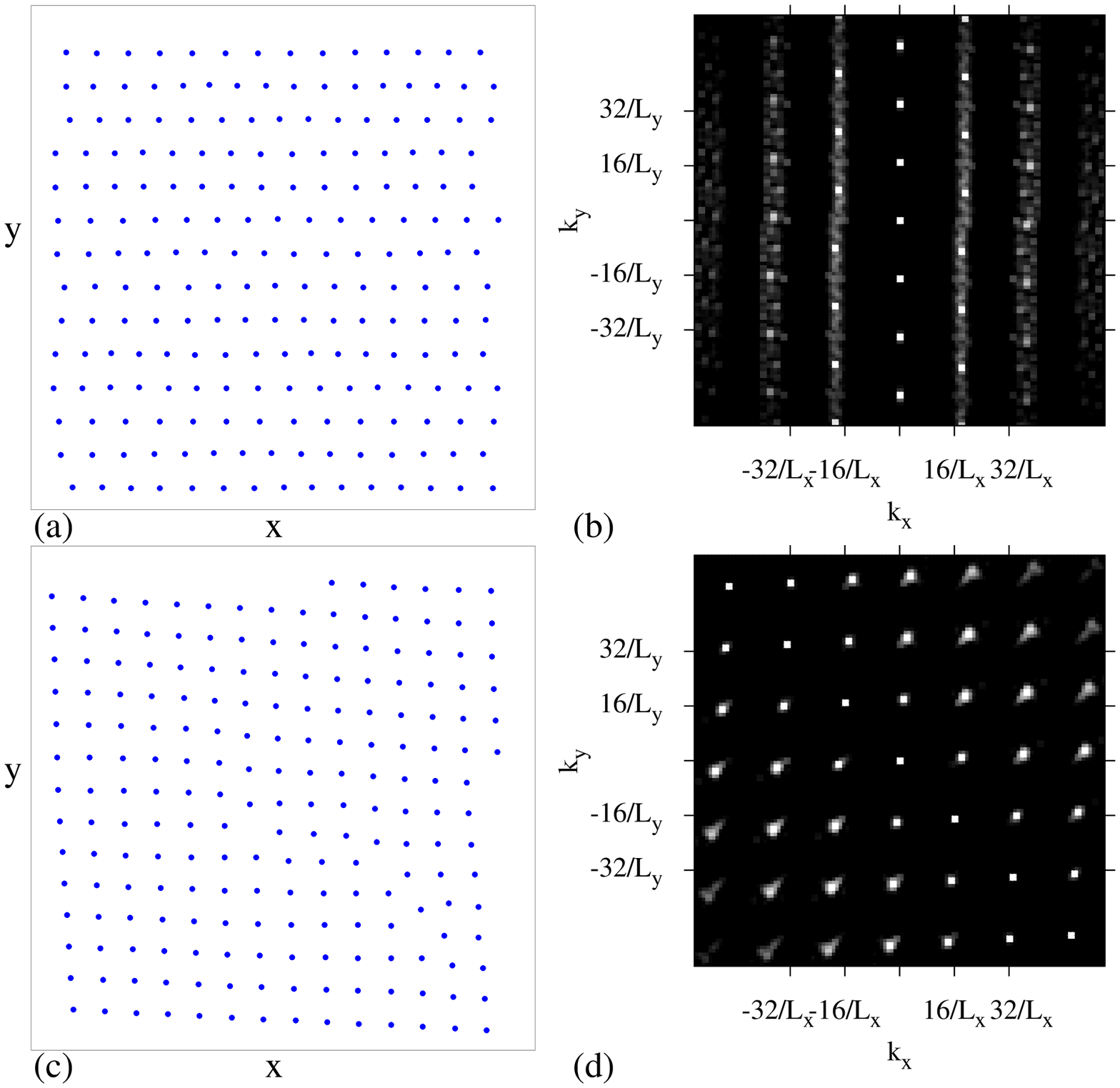}
\caption{
The vortex positions (a,c) and corresponding $S(k)$ (b,d) for the
square pinning system from figure 10 with $F_p=1.5$ at 
$B/B_{\phi} = 1.5$. (a,b) The $(0,1)$ step. (c,d) The $(1,2)$ step.  
\label{Fig11}
}
\end{figure}

\section{Square Pinning Array}

The same general behaviour found for the triangular pinning array also occurs
for vortices moving over square pinning arrays.
In figure~10(a) we plot $V_{y}$ versus drive angle $\theta$ 
for a sample containing a square pinning array with $F_{p} = 1.5$ 
for $B/B_{\phi} = 0.2$, 1.5, and $4.0$.
In the square array, locking steps appear when $\theta = \tan^{-1}(m/n)$ 
with $m$ and
$n$ integers, and in the figure we highlight steps with $(m,n)$= 
(0,1), (1,3), (1,2), (1,1), (2,1), (3,1), and (1,0). 
The most prominent steps fall at (0,1) for $\theta=0^\circ$ and (1,1) for
$\theta=45^\circ$.
At $B/B_{\phi} = 4.0$, many of the higher order steps are very 
small or or missing and the (1,1) step is much reduced in size while
the (1,2) and (2,1) steps remain prominent. 
In figure~10(b) we plot the width of the (0,1) step 
$F_{c}^{Tr}$ vs $B/B_{\phi}$.  The behaviour of $F_c^{Tr}$ is very similar to that shown
for the triangular pinning array in figure 6.  For the square array,
local minima in $F_c^{Tr}$ appear at $B/B_{\phi} = 0.6$, 2.0, and $5.5$. 
At the local maxima of $F_c^{Tr}$, we find the same one-dimensional channeling
effects described in the previous section for the triangular pinning arrays.
The structure of the vortex lattice on the steps in the square pinning
array system is also similar to that
found for the triangular arrays, as illustrated in figure 11.
For example, a smectic structure appears at 
$B/B_\phi=1.5$
on the (0,1) step, shown in figure 11(a,b).
One distinction is that the square pinning array system shows a much larger
number of locking steps where the vortices form a square moving lattice
structure, such as that illustrated for the (1,2) step at $B/B_\phi=1.5$
in figure 11(c,d).

\section{Substrate Strength and Different Locking Regimes} 

We next consider the effect of substrate strength on the locking regimes
for a system with a triangular pinning lattice at fixed $B/B_\phi=1.11$.
In figure~12 we plot $V_{y}$ versus $\theta$ for samples with
$F_{p} = 0.25$, 0.5, 1.0, and 1.5.
The step widths grow with increasing $F_{p}$. 
A similar effect was observed for locking on square pinning 
arrays in previous work \cite{Nori}. 
As the substrate strength becomes very large,
we observe a new phenomenon where the locking effects become reduced. 
The plot of $V_y$ versus $\theta$ for $F_p=2.35$ in figure~13(a) 
shows that although the $(0,0)$ and the $(1,0)$ locking steps are
present, all of the remaining steps are gone and are replaced
by a strongly fluctuating regime. 
Some directional guidance effects still occur within this strongly
fluctuating regime, as shown by the dip in $V_y$ appearing at
$\theta=90^\circ$ and 
the shoulder feature near $\theta = 36^{\circ}$. 
In the non-step regions, 
the flow is strongly disordered.  The trajectories of the particles do
not fall into ordered channels but mix strongly, and vortices become
pinned and depinned at random.
Although ordered flow occurs along the $(0,0)$ and $(1,0)$ steps, 
it differs from the ordered flow found at the lower pinning forces 
shown in figure~12. 
We find that collective effects between the particles begin to dominate
the behaviour once
$F_{p} > 2.0$.
In a system with a vortex density low enough to fall into the single
particle behaviour limit, all of the vortices are immobilized for
$F_p>2.0$ since the pinning strength exceeds the magnitude of the driving
force, $F_p>A$.
For all of the values of $F_p$ shown in figure 12, we have $F_p<A$ and there
are no pinned vortices; instead, all of the vortices flow along or between
the pinning sites.
On the locking steps at 
$F_{p} = 2.35$, motion occurs in the form of a pulse of depinned vortices
which passes through a background of pinned vortices.  Each vortex spends
part of the time pinned and part of the time moving, and there are always
some pinned vortices present, but the identity of the the pinned vortices is
constantly changing.
This is why $V_{y}$ along the 
$(1,0)$ 
step for $F_p=2.35$ is nearly a factor of 3 lower than $V_y$ along the
(1,0) step for $F_p=1.5$, where the vortices move continuously and are never
pinned.
A similar soliton or incommensurate flow of vortices 
along rows of pinning sites has been observed previously
for vortices driven along a single principle axis of periodic pinning
arrays \cite{Commensurate}.  Additionally, 
a transition from a higher velocity random or turbulent type of vortex flow 
to a lower velocity flow state with smaller fluctuations was also observed
in Ref.~\cite{Commensurate}, a feature which resembles the transition
in and out of the (1,0) step in figure 13(a).
We find a very pronounced jump up in $V_{y}$ at the end of each of the
two locking steps
which is correlated with a
sudden jump in $V_{x}$ (not shown).  

\begin{figure}
\includegraphics[width=3.5in]{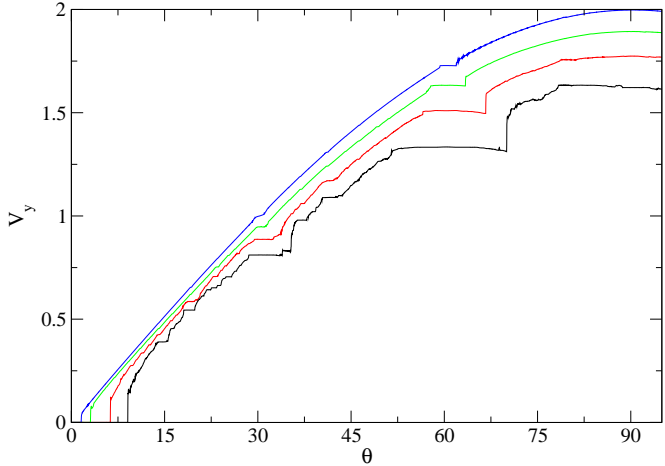}
\caption{
$V_{y}$ vs $\theta$ for a triangular pinning array at $B/B_{\phi} = 1.11$ for
different substrate strengths
$F_{p} = 0.25$, 0.5, 1.0, and 1.5, from top to bottom.
Here the step widths grow with increasing $F_{p}$.
\label{Fig12}
}
\end{figure}

Figure 13(b) shows that for $F_{p} = 2.5$, all the steps are gone and the
flow is always in the strongly fluctuating regime.
The former steps are replaced with local minima in $V_y$ at the 
angles where the $(0,1)$, $(1,1)$, $(1,0)$, and $\theta=90^{\circ}$ 
locking steps would have occured, indicating that there is 
still a guidance effect from the substrate but not complete locking
of the vortex motion. 
At $F_{p} = 2.75$ in figure 13(c), 
the (0,0), (1,1), and $\theta=90^{\circ}$ locking steps reappear and 
a dip in $V_y$ persists at $\theta=60^{\circ}$ where the (1,0) step would
have been.  For $F_{p} = 2.75$, figure 13(d) shows that the
(0,1), (1,2), (1,1), (2,1), and $\theta=90^{\circ}$ steps are now all restored, 
while between the steps the flow enters a rapidly fluctuating phase in
which $V_y$ is significantly enhanced.
In figure 13(d), along a given locking step
$V_y$ is not constant but 
has a curved shape and decreases noticeably just before
the end of each step.
The vortex flow on the steps occurs in the form of an incommensurate or 
solitonlike pulse of depinned vortices which passes through a background of
pinned vortices.
For $F_{p} > 2.9$, the vortices transition directly from one locking step
to the next locking step and there are
no longer regions of random flow between the steps, 
as illustrated in figure 14 for a sample
with $F_p=3.0$ that exhibits a rich variety of locking steps.
As $F_{p}$ increases further, the $V_{y}$ versus $\theta$ curves retain the
shape shown in figure 14 
with small shifts in the steps until $F_{p} \ge 3.9$, 
at which point all the vortices become pinned for all $\theta$.        

\begin{figure}
\includegraphics[width=3.5in]{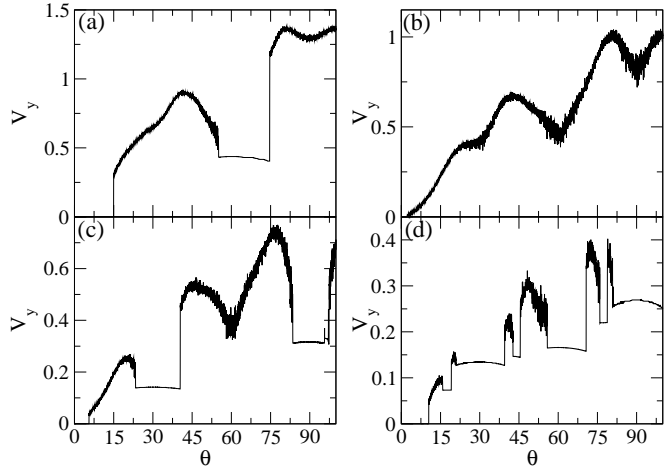}
\caption{
$V_{y}$ vs $\theta$ for the sample from figure 12 with a triangular pinning
array at $B/B_{\phi}  = 1.11$.
(a) $F_{p} = 2.35$. There are only two locking regions separated by
strongly fluctuating regions where some guided motion of the particles
occurs.
(b) $F_{p} = 2.5$.  The locking regions are almost completely absent.
(c) $F_{p} = 2.75$.  Several locking regions reappear. 
(d) $F_{p} = 2.85$.  There are a larger number of locking steps 
intermixed with randomly fluctuating regions.  
\label{Fig13}
}
\end{figure}

\begin{figure}
\includegraphics[width=3.5in]{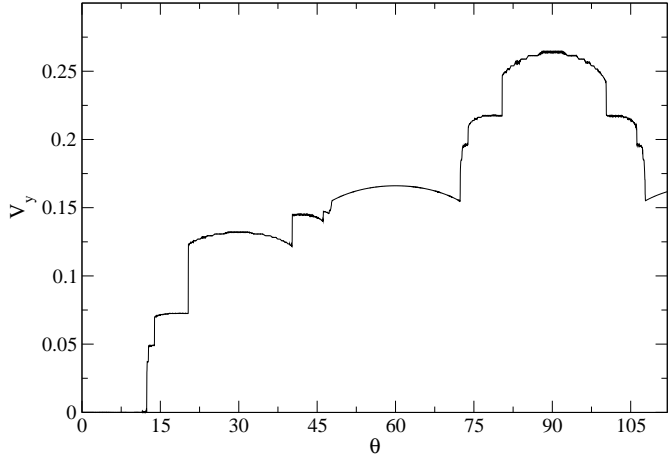}
\caption{
$V_{y}$ vs $\theta$ for a triangular pinning array at $B/B_{\phi} = 1.11$ 
with $F_{p} = 3.0$. 
The random fluctuating regimes are lost and the system passes directly
from one locking phase to the next.
\label{Fig14}
}
\end{figure}

\begin{figure}
\includegraphics[width=3.5in]{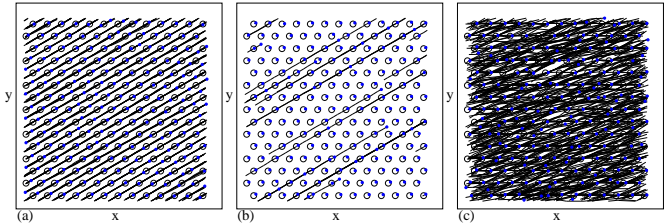}
\caption{
The positions of the pinning sites (large open circles) and vortices (small
filled circles) as well as vortex trajectories over a fixed time interval
(lines) in a portion of a sample with a triangular pinning array
on the (1,1) locking step at $\theta=30^\circ$.
(a) In the weak pinning regime at $F_{p} = 1.75$, an ordered flow
occurs with all the vortices moving along the $\theta=30^\circ$ direction.
(b) In the strong pinning regime at $F_{p} = 2.75$, a portion of the vortices
are pinned and the flow occurs by a pulse motion or flowing kink. 
(c) At $F_{p} = 2.75$ for $\theta = 15^\circ$, 
a non-locking fluctuating flow phase occurs.  
\label{Fig15}
}
\end{figure}

\begin{figure}
\includegraphics[width=3.5in]{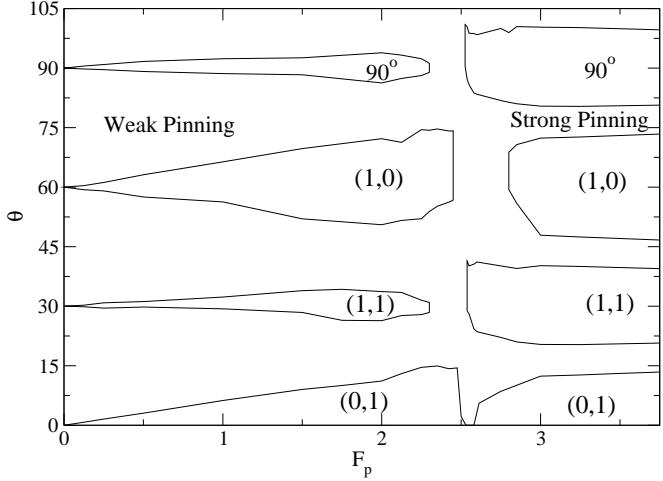}
\caption{
The dynamic phase diagram of $\theta$ vs $F_{p}$ for a sample with
a triangular pinning array at $B/B_{\phi} = 1.11$. 
The regions in which the (0,1), (1,1), (1,0), and $90^{\circ}$ locking steps
appear are marked.
The left side of the diagram at lower $F_p$ is the weak pinning
regime.
Near $F_{p} = 2.5$ a large portion
of the vortex dynamics falls in the random fluctuating phase rather than
on locking steps.
In the strong pinning regime the second type of locking steps appear. 
For $F_{p} \ge 3.9$ all the vortices are pinned.   
\label{Fig16}
}
\end{figure}

To demonstrate the differences in the flow for samples with strong and weak 
pinning, in figure~15(a,b) we plot the vortex trajectories 
on the (1,1) locking step at $\theta=30^\circ$ over a fixed 
time interval. 
For $F_{p} = 1.75$, figure 15(a) 
shows that all the vortices move along one-dimensional paths 
oriented along the $\theta=30^\circ$ driving direction.
For $F_{p} = 2.75$ in figure 15(b), only a portion of the vortices 
move in one-dimensional paths along the pinning rows while the remaining
vortices are pinned.
Figure~15(c) shows that in the $F_{p} = 2.75$ sample at $\theta = 15^\circ$,
a fluctuating flow phase occurs and there is no longer any one-dimensional
channeling of the vortex motion.
For samples with strong substrates, along the locking steps the vortices
generally exhibit a triangular structure since a large portion of the vortices
are trapped by the pinning sites, so the large scale structural transitions
found for samples with weak substrates are lost.
By conducting a series of simulations for varied $F_{p}$ we map 
the transition between the strong and weak substrate regimes, as shown
in figure 16 where we highlight the widths of the
(0,1), (1,1), (1,0), and $\theta=90^\circ$ locking steps.
The (0,1) step increases in width with increasing $F_p$ 
up to $F_{p} = 2.45$, after which the step vanishes. 
This denotes the transition
from the weak pinning locking regime to the random fluctuating phase. 
For $F_{p} > 2.55$ the (0,1) step reappears in 
the strong pinning regime and its width saturates for $F_{p} > 3.0$.
The higher order steps also show similar features with the step widths 
diminishing to zero at $F_{p} = 0.0$ and the steps vanishing
close to $F_{p} = 2.5$. 
There are also many other higher order steps not shown in 
figure~16 which in general have similar behaviour.

\begin{figure}
\includegraphics[width=3.5in]{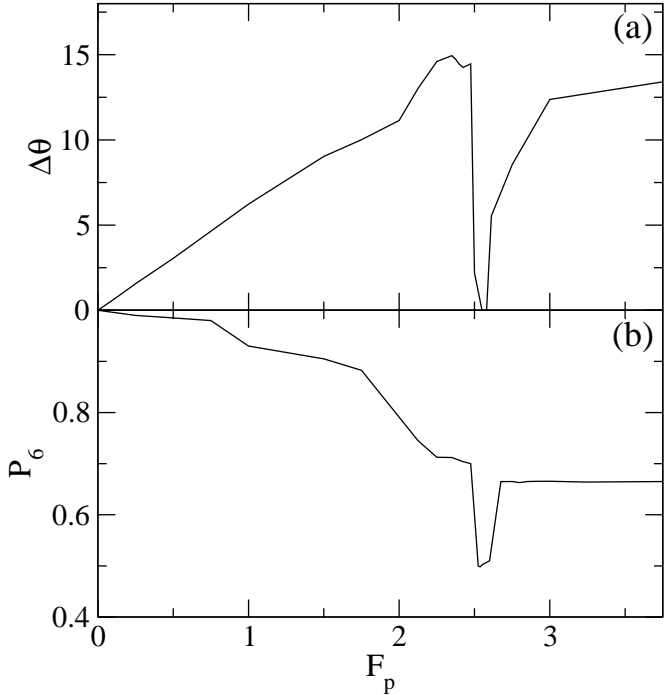}
\caption{
(a) The width $\Delta\theta$ 
of the $(1,0)$ locking phase from figure 16 vs $F_{p}$.
(b) The corresponding fraction of six-fold coordinated particles 
$P_{6}$ has a dip in the random fluctuating phase 
which separates the weak and strong pinning regimes.  
\label{Fig17}
}
\end{figure}

\begin{figure}
\includegraphics[width=3.5in]{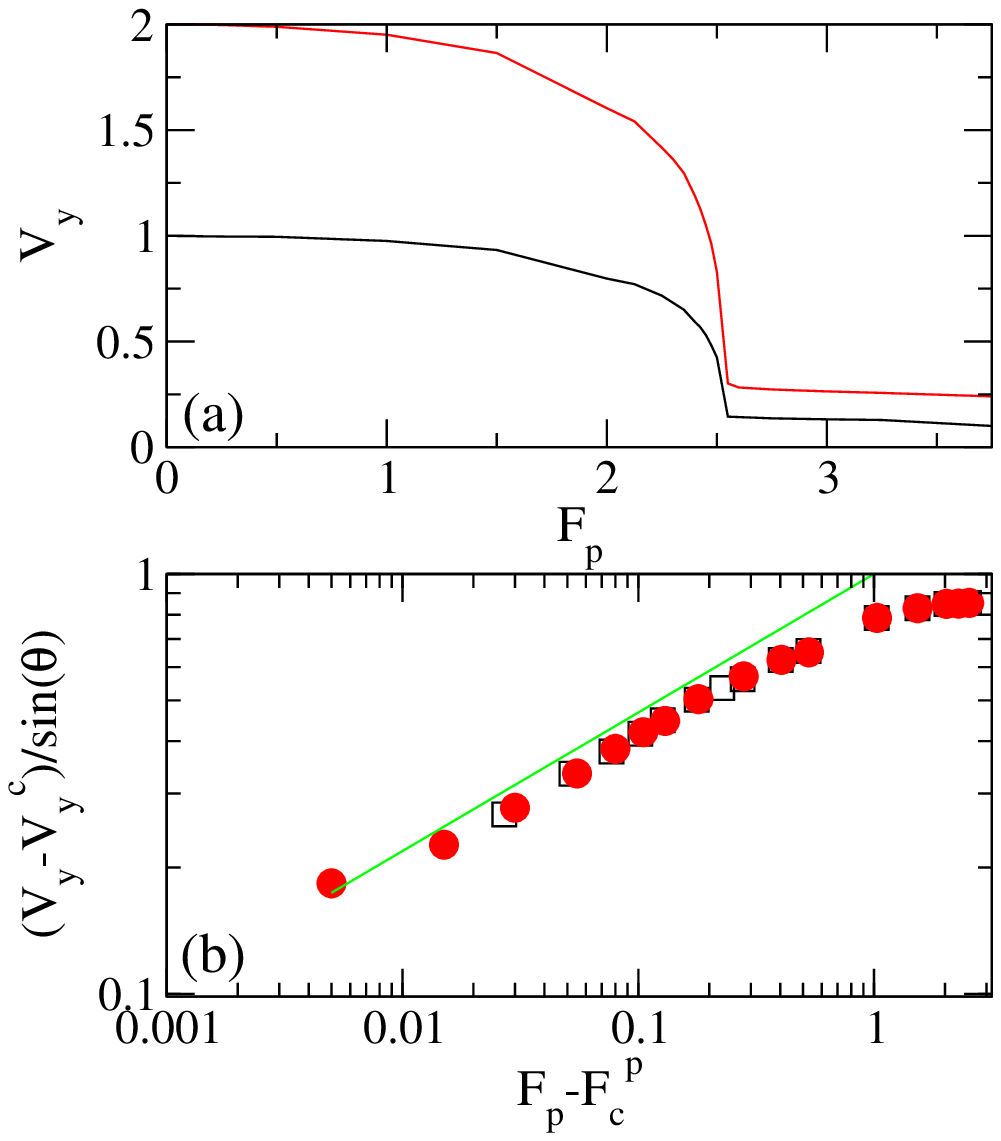}
\caption{
(a) The average velocity $V_y$ vs $F_{p}$ for the system in figure 16 
at $\theta = 90^{\circ}$ (upper curve) and $\theta = 30^{\circ}$ 
(lower curve) showing that
in the strong pinning regime the particle velocity saturates.
(b) A scaling collapse of the same curves near the crossover 
at $F_c^p$ from
the weak pinning to the strong pinning regime.
The line is a fit to $V_{y} - V^{c}_{y} \propto (F_{p} - F_{c}^p)^{\alpha}$  with
$\alpha=1/3$, where
$V^{c}_{y}$ is the value of the $V_y$ in the strong pinning regime.
\label{Fig18}
}
\end{figure}

In figure~17 we show that the crossover from the weak pinning to the strong
pinning regime can also be identified by measuring $P_6$ versus
$F_p$.  We plot the width $\Delta\theta$ of the (1,0) 
step from figure 16 in figure 17(a),
and show the corresponding $P_6$ in figure 17(b).
At $F_{p} = 0.0$, $P_{6} = 1.0$,
and $P_6$ generally decreases with increasing $F_p$ 
as more dislocations enter the system
in the moving smectic phase until reaching $P_{6} = 0.7$ 
at $F_{p} =2.4$. There is then a dip in $P_{6}$ at $F_p=2.5$
where the step width $\Delta\theta$ drops to zero.
For $F_p>2.5$,  $P_{6}$ increases with increasing $F_p$ until reaching 
$P_6=0.67$ when the system enters the strongly pinned regime 
and a large portion of the vortices become pinned 
in the triangular substrate, increasing the amount of sixfold ordering 
present.   
It is also possible to detect the crossover between the weak and strong
pinning regimes using
the average velocity in the $y$-direction  
$V_{y}$ versus $F_{p}$ for a given $\theta$, 
as shown in Figure~18(a) for $\theta = 90^\circ$  and 
$\theta = 30^\circ$. 
At $F_{p} = 0.0$ the values of the velocities are simply 
$V_y=A\sin(30^\circ) = 1.0$ and 
$V_y=A\sin(90^\circ) = A = 2.0$. 
As $F_{p}$ increases, $V_{y}$ monotonically 
decreases until just above $F_{p} = 2.5$ which marks the 
crossover to the strong pinning regime in which
$V_{y}$ remains constant for increasing $F_{p}$. 
The decrease in $V_y$ in the weak pinning regime can be fit to the
functional form
$V_{y} - V^{c}_{y} \propto  (F_{p} - F^p_{c})^{\alpha}$, 
where $V^{c}_{y}$ is the saturation value of the velocity
in the strong pinning regime and 
$F^p_{c}$ is the critical value of the pinning force
at which the system enters the strong pinning regime, 
identified as the value of $F_p$ above which $V_y$ becomes constant.
In figure~18(b) 
we show this scaling 
with $V_y$ normalized by $\sin(\theta)$.
Here the $\theta=30^\circ$ and $\theta=90^\circ$ curves
collapse on each other and the solid line indicates a fit with $\alpha = 1/3$.  
The behaviour of the $V_{y}$ curves varies on different sets of
locking steps, 
so a straightforward scaling such as that shown in figure~18(b) 
is not always possible.  For example, in figure~19 we plot
$V_{y}$ versus $F_{p}$ on the $(1,0)$ step as well as for a 
driving angle of $\theta=17^\circ$. 
In both cases $V_{y}$ decreases
with increasing $F_{p}$ for $F_{p} < 2.5$. 
For 
$\theta = 17^\circ$, 
$V_y$ drops to zero within the weak pinning regime
when the (0,1) step widens to include this driving angle and
the vortices move in one-dimensional channels that are aligned
with the $x$ axis.
Near $F_{p} = 2.5$, $V_y$ for both driving angles 
passes through a local maximum when the system 
enters the random fluctuating flow regime.
At $F_{p} = 2.85$ for $\theta=60^\circ$, the $(1,0)$ step 
reappears in the strong pinning regime, where $V_{y}$ remains
constant.  Similarly for $\theta=17^\circ$ $V_y$ saturates to
a nearly constant value in the strong pinning regime.

\begin{figure}
\includegraphics[width=3.5in]{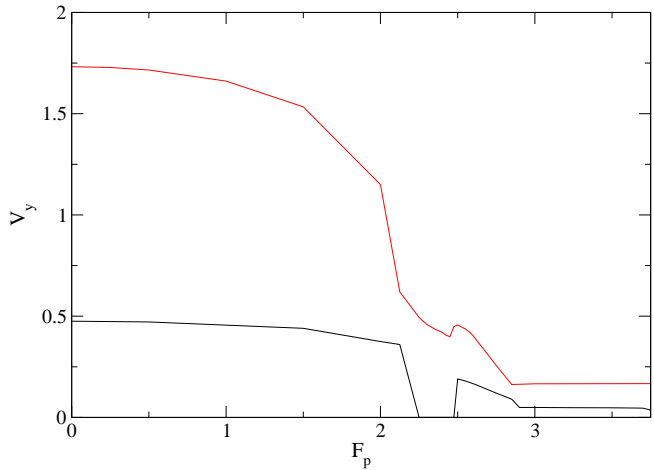}
\caption{
The average velocity $V_y$ vs $F_{p}$ for the system in 
figure~16 at $\theta = 60^\circ$ on the $(1,0)$ step (upper curve) 
and at $\theta = 17^\circ$ (lower curve).
Here a local maximum in $V_y$ occurs in the random fluctuating regime
which separates the weak and strong pinning regimes. 
\label{Fig19}
}
\end{figure}

\begin{figure}
\includegraphics[width=3.5in]{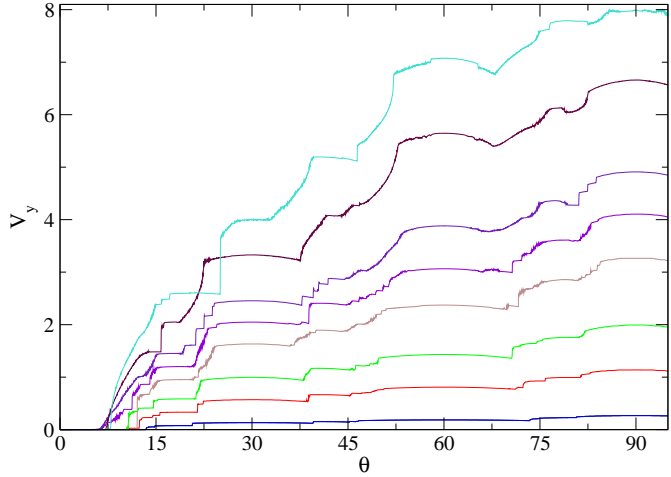}
\caption{
$V_{y}$ vs $\theta$ for a triangular pinning array with 
$F_{p} = 3.75$ at 
$B/B_{\phi} = 1.11$, 1.48, 1.852, 2.407, 2.78, 3.148, 4.0,
 and $4.78$, from bottom to top.  
\label{Fig20}
}
\end{figure}

\begin{figure}
\includegraphics[width=3.5in]{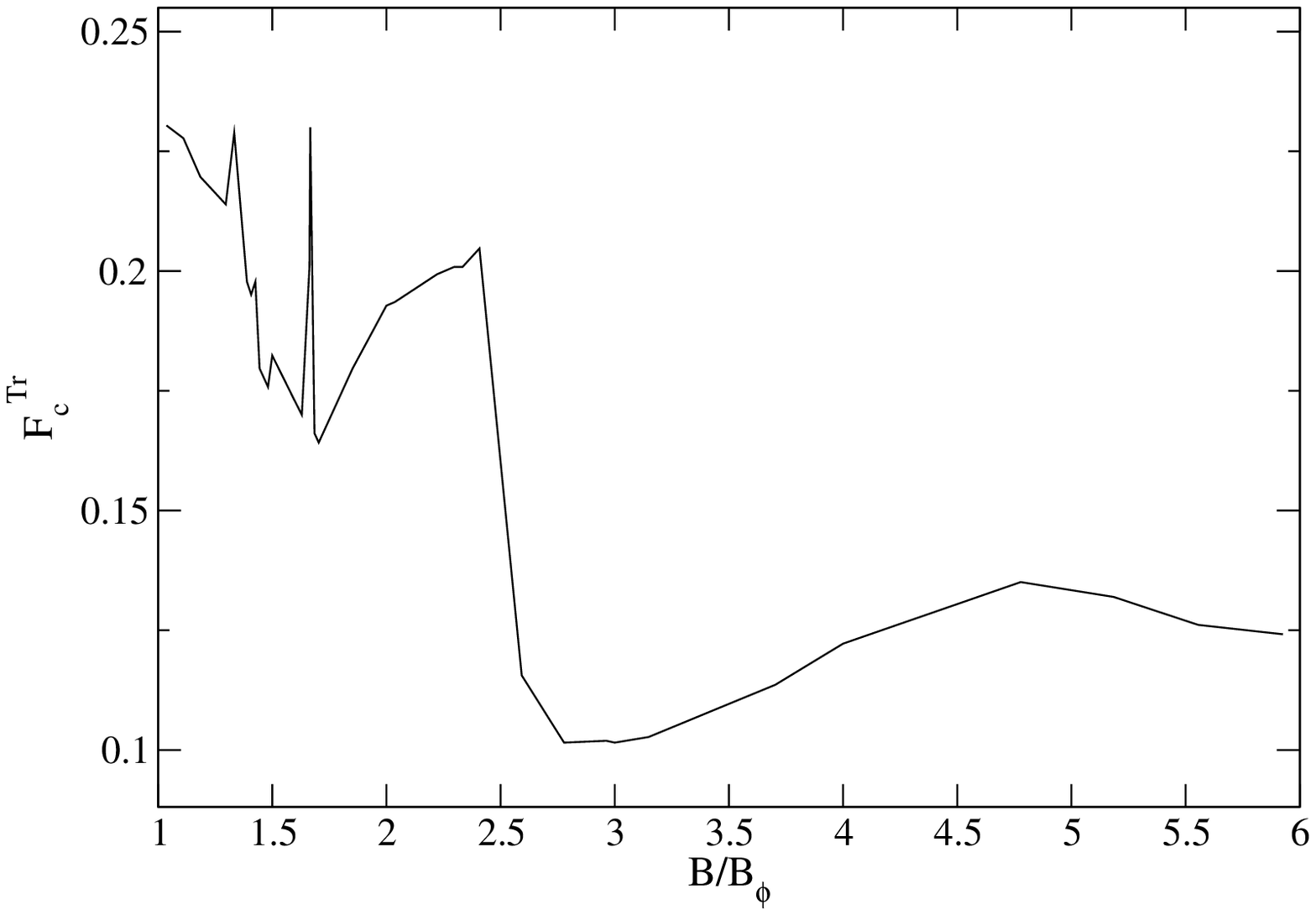}
\caption{
The width $F_c^{Tr}$ of the (0,0) step expressed in terms of the $y$ component
of the driving force at the end of the step
vs $B/B_{\phi}$ for the triangular pinning system from figure 20.  
\label{Fig21}
}
\end{figure}

We next consider the effects of changing 
$B/B_{\phi}$  on the behaviour in the strong pinning regime. 
In figure 20 we plot $V_{y}$ versus $\theta$ for a 
triangular pinning system with $F_{p} = 3.75$
at $B/B_{\phi} = 1.11$, 1.48, 1.852, 2.407, 2.78, 3.148, 4.0, and $4.78$.
The system is pinned at all $\theta$ for $B/B_{\phi} < 1.0$.    
For all of the fillings shown in figure 20,
strong locking effects occur, and the widths of some of the locking steps
vary with $B/B_\phi$.
For $B/B_{\phi} > 2.78$ the (1,0) locking step 
shows increased curvature 
and regions of negative differential conductivity appear
near $\theta = 38^\circ$, $68^\circ$, and $ 78^\circ$, while
there is a dome like feature on the $\theta=90^\circ$ locking step.  
At some values of $B/B_\phi$,
a series of smaller locking steps appear, but these small steps disappear for
larger values of $B/B_\phi$.
In general, all of the smaller steps are washed out for 
increasing $B/B_{\phi}$. 

In figure~21 we plot the width $F_c^{Tr}$ of the (0,0) step 
versus $B/B_{\phi}$ for the strong pinning regime sample from
figure 20. 
The broad features of the curve are similar to the behaviour of $F_c^{Tr}$
in the weak pinning regime that was shown in figure 6.
There are several local minima and maxima in $F_c^{Tr}$ and
a broad maximum in the range $ 3.5 < B/B_{\phi} < 5.5$,
as well as another local maximum near $B/B_{\phi} = 2.4$. 
As in the weak pinning regime, for the strong pinning regime the
maxima in $F_c^{Tr}$
are not centred at integer matching fields. 
Unlike the weak pinning regime, however, there are no fillings at which
$F_c^{Tr}$ drops to zero.  We also observe sharp and narrow peaks at
$B/B_{\phi} = 4/3$ and  $B/B_\phi=5/3$, while no comparably sharp peaks in
$F_c^{Tr}$ occured in the weak pinning regime.
Longitudinal depinning studies performed for vortices moving over
triangular pinning arrays have demonstrated enhanced pinning
at fractional matching fields
such as $B/B_\phi=1/3$,  $B/B_\phi=2/3$, and higher order multiples
where the vortices can form submatching configurations with triangular
ordering \cite{Jensen}.
The vortex flow on the (0,0) step for $B/B_{\phi} < 2.5$ occurs
via the same flow of incommensurations along the pinning sites that was
illustrated in figure 15(b),
and the peaks in $F_c^{Tr}$ at $B/B_\phi=4/3$ and 
$B/B_\phi=5/3$ correspond to fillings at which the
moving incommensurations can form a triangular ordering. 
We expect that for a square pinning lattice in the strong pinning
regime, peaks in the width of the first step 
would occur at $B/B_\phi=3/2$, $5/4$, and $7/4$. 
In figure 21 for $B/B_{\phi} > 2.5$, 
a portion of the vortices begin to flow in the interstitial regions,
which may prevent the formation of higher order fractional
matching peaks.    

\begin{figure}
\includegraphics[width=3.5in]{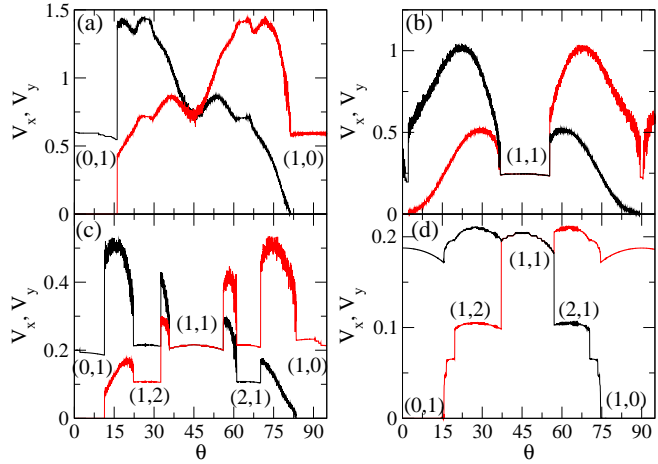}
\caption{
$V_{x}$ (dark line) and $V_{y}$ (light lines) vs $\theta$ for 
a sample with a square pinning array at $B/B_{\phi} = 1.11$. 
(a) At $F_{p} = 2.25$, only two locking steps are present which are
separated by a random fluctuating phase.
(b) At $F_{p} = 2.5$ the $(1,1)$ locking step appears. 
(c) At $F_{p} = 2.75$ we find a mixture of locking steps and 
randomly fluctuating phases. 
(d) At $F_{p} = 3.75$ the system transitions directly from one locking
step to another with no randomly fluctuating regimes.
\label{Fig22}
}
\end{figure}

We find the same transition from a weak pinning regime to a strong pinning
regime for square pinning arrays.  In figure 22(a) 
we plot $V_{x}$ and $V_{y}$ versus $\theta$
for a sample with square pinning at $F_{p} = 2.25$ 
and $B/B_{\phi}  = 1.11$, where
locking steps and randomly fluctuating phases appear.
On the (0,1) and (1,0) locking steps, the vortices all flow along rows or
columns of pinning sites.  At $\theta=45^\circ$ where the (1,1) step
should appear, we find no step but instead both $V_x$ and $V_y$ pass through
a local minimum.
There are also some weak locking shoulder
features in $V_{x}$ and $V_{y}$ near 
$\theta=26^\circ$ and $\theta=65^\circ$. 
Even though $F_y$ reaches its maximum cycle value at $\theta=90^\circ$, 
$V_y$ sits in a local minimum at this driving angle.
For $F_{p} = 2.5$ shown in figure 22(b), the $(0,1)$ and $(1,0)$ 
steps have almost completely disappeared;
however, there is now a clear (1,1) locking step.
At $F_p=2.75$, figure~22(c) indicates that locking steps emerge
at (0,1), (1,2), (1,1), (2,1), and $(0,1)$, 
with large fluctuating regions falling between adjacent
locking steps. 
For $F_{p} > 3.0$ the system enters the strong pinning regime and 
the random fluctuating regions disappear, as shown in 
figure~22(d) for $F_{p} = 3.75$.
Here, $V_y$ is higher on the (2,1) step than on the (1,0) step even though
no random fluctuating region separates the two steps.
The length of the moving incommensurations 
on the $(2,1)$ step is slightly larger than the   
length of the moving incommensurations along the $(1,0)$ step, leading to
the higher value of $V_y$ for the (2,1) step. 

\begin{figure}
\includegraphics[width=3.5in]{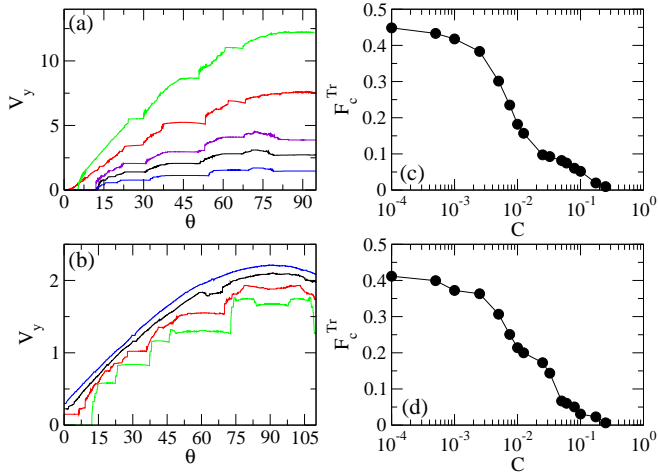}
\caption{
\label{Fig23}
(a) $V_{y}$ vs $\theta$ for colloidal particles 
with interaction strength $C=0.001$ on a 
square pinning array with $F_p=1.85$.
The ratio of the number of colloids to pinning sites is 
$N_{c}/N_{p} =0.432$, 0.778, 1.125, $1.82$, and 2.85, from bottom to top. 
(b) $V_{y}$ vs $\theta$ for colloidal 
particles on a triangular pinning array with $F_{p} = 1.85$ for
$N_{c}/N_{p} = 0.432$ 
and varied colloid-colloid interaction strength 
of $C = 0.0001$, 0.01, 0.05, and $0.1$, from bottom to top. 
(c) The width $F_c^{Tr}$ of the first locking step from the 
square pinning array sample in panel (a) at $N_c/N_p=0.432$
vs $C$. (d) The same as (c) for a triangular pinning array   
with the same parameters as in panel (b). 
}
\end{figure}

\section{Colloidal Particles Moving Over Triangular and Square substrates}
Another question is how general the results we have obtained for the
vortex system are
for other types of particle interactions, 
particularly colloidal particles where 
there is currently considerable interest in kinetic locking effects. 
We have tested all of the major predictions 
for the vortex system on  square and triangular
pinning lattices for colloidal particles interacting with a screened 
Yukawa potential.
In figure~23(a) we plot $V_{y}$ vs $\theta$ for colloidal
particles on a square pinning array of strength $F_p=1.85$
for varied colloid densities of $N_{c}/N_{p} = 0.432$, 0.778, 1.125, 1.82, and 
$2.85$.  The colloid-colloid interaction coefficient is $C= 0.001$.   
Here we find the same
locking step features observed for the vortex system, with the same oscillations
in the width of the first and higher order steps.
At $N_{c}/N_{p} = 1.82$ the width of the first step drops to zero, but 
the first step reappears at $N_{c}/N_{p} = 2.85$,
similar to what was observed for the vortex system. 
We also find the same behaviours for colloids on triangular
pinning arrays.  
One quantity that can be varied in the colloidal system but 
not in the vortex system
is the particle-particle interaction strength. 
This can be achieved either by increasing the
effective charge of the colloids or by changing the screening length 
while keeping the strength of the substrate
fixed. In figure~23(b) we plot $V_{y}$ versus $\theta$ 
for colloids on a triangular pinning lattice  at fixed 
$F_{p} = 1.85$ and $N_{c}/N_{p} = 0.432$ with 
$C=0.0001$, 0.01, 0.05, and  $0.1$. 
As $C$ increases, the widths of the locking steps decrease since the
colloid
lattice becomes stiffer and is less able to distort to adjust to the pinning. 

Figure~23(d) shows the width $F_c^{Tr}$ of the first locking step as given by
the value of the $y$ component of the driving force at the end of the step
as a function of $C$ for colloids on a triangular pinning lattice from the
system in figure 23(b).
As $C$ increases, $F_c^{Tr}$ gradually decreases to zero.
There is no clear transition between weak and strong pinning regimes when
$C$ is varied, unlike the transition found when $F_p$ was varied for the
vortex system.
This is because our choice of fixed $F_{p} = 1.85$ 
falls below the driving amplitude of $A = 2.0$, so all the 
colloids are moving even when the particle-particle interaction strength
is negligible.
In order to observe the strong pinning
regime discussed earlier, it is necessary for 
a portion of the particles to be pinned and for the motion to occur in the
form of depinned incommensurations which pass through the system.
In figure~23(c) we plot the width $F_c^{Tr}$ of the first locking step versus 
$C$ for colloids on a square pinning array with the same parameters used
in figure 23(a),
indicating that the same behaviour found for colloids on
a triangular pinning lattice also occurs for
colloids on a square pinning lattice.  

\section{Summary} 

In summary, we have investigated the collective ordering and
disordering effects on directional locking for
particles such as vortices and colloids moving over 
triangular and square substrate arrays.  We identify
several different regimes of collective behaviour
as a function of the substrate strength and of the ratio of particle density
to substrate minima density. 
For weak substrates, at certain driving angles all the particles flow 
along one-dimensional channels through the pinning sites,
generating a series of constant velocity steps on which the motion 
remains locked to a certain direction over a range of driving angles. 
As the pinning strength 
decreases, the width of the locking steps decreases and 
larger ranges of driving angle are occupied by nonlocking regions
in which the particle-particle interactions dominate and a triangular particle
lattice forms.
On the locking steps for triangular substrate arrays,
we find that a rich variety of different types of moving 
lattices form, including moving smectic, square, anisotropic disordered, 
and distorted triangular lattice structures. 
The most prominent locking steps are associated with smectic type 
particle orderings. 
In the weak substrate regime,
we observe that the widths of the steps including the 
initial transverse depinning barrier pass through local minima
and maxima as a function of the ratio of 
particle density to substrate minima density. 
In contrast to the longitudinal depinning threshold for periodic substrates,
which shows peaks at commensurate fields, 
the local maxima for the width of the first locking step 
are not correlated with the two-dimensional periodicity of 
the pinning array but are instead related to 
a dynamical commensuration effect caused by the formation of one-dimensional
channels of moving particles.
A local maximum in the width of the first locking step 
occurs when integer numbers of rows of moving vortices can fit
in the interstitial areas between the pinning sites. 
For fillings at which the moving rows are unable to fit without 
buckling, the particle structure becomes disordered and the
width of the first locking step is small or zero.
As a function of substrate strength for fixed
particle density, we identify two distinct locking regimes: 
a weak pinning regime where all the particles flow along the pinning sites, 
and a strong pinning regime where the flow occurs by means of 
an incommensuration or a pulse passing through a background of pinned 
particles. Between these
two regimes the locking steps are lost and are
replaced by a strongly fluctuating regime where the particle motion does not
lock to a particular direction. 
For a fixed driving angle, 
the average particle velocity drops sharply at the 
crossover between these two regimes, and the velocity saturates to a plateau
value in the strong substrate limit.
In the strong substrate regime
the width of the first step displays commensurate peaks when the number 
of particles is a fractional matching ratio of the number of substrate
minima, in addition to showing peaks at the incommensurate fields as in
the weak pinning regime.
We expect these effects to be relevant to a wide class of collectively 
interacting particles moving over periodic 
substrates, such as vortices in type II superconductors, 
colloids, electron crystals, 
and other soft matter systems.

\section{Acknowledgments}
This work was supported by the U.S.~Department of Energy under
Contract No.~W-7405-ENG-36.

\end{document}